\documentstyle[aps,prd,amstex,epsfig]{revtex}

\begin{document}

\draft
\author{Christophe Ringeval}
\address{D\'epartement d'Astrophysique Relativiste et de Cosmologie,\\
Observatoire de Paris-Meudon, UMR 8629, CNRS, 92195 Meudon, France,\\
Institut d'Astrophysique de Paris, 98bis boulevard Arago, 75014 Paris,
France.}
\title{Equation of state of cosmic strings with fermionic current-carriers}
\date{\today}
\maketitle
\newcommand{\ETAL}{{\it et al.}}
\newcommand{\ud}{{\mathrm{d}}}
\newcommand{\ue}{{\mathrm{e}}}
\newcommand{\uV}{{\mathrm{V}}}
\newcommand{\Chi}{{\mathcal{X}}}
\newcommand{\Rho}{{\mathcal{R}}}
\newcommand{\Rhop}{\widehat{{\mathcal{R}}}}
\newcommand{\cpsir}{c_{\psi_R}}
\newcommand{\cpsil}{c_{\psi_L}}
\newcommand{\cchir}{c_{\chi_R}}
\newcommand{\cchil}{c_{\chi_L}}
\newcommand{\cphi}{c_{\phi}}
\newcommand{\chip}{{\chi_{\mathcal{P}}}}
\newcommand{\psip}{{\psi_{\mathcal{P}}}}
\newcommand{\jbt}{\tilde{\overline{j}}}
\newcommand{\I}{I}

\begin{abstract}
The relevant characteristic features, including energy per unit length
and tension, of a cosmic string carrying massless fermionic currents
in the framework of the Witten model in the neutral limit are derived
through quantization of the spinor fields along the string. The
construction of a Fock space is performed by means of a separation
between longitudinal modes and the so-called transverse zero energy
solutions of the Dirac equation in the vortex. As a result,
quantization leads to a set of naturally defined state parameters
which are the number densities of particles and anti-particles trapped
in the cosmic string. It is seen that the usual one-parameter
formalism for describing the macroscopic dynamics of current-carrying
vortices is not sufficient in the case of fermionic carriers.
\end{abstract}


\section{Introduction}

The mechanism of spontaneous symmetry breaking involved in early
universe phase transitions in some Grand Unified Theories (GUT)
might lead to the formation of topological defects~\cite{kibble}.
Among them, only cosmic strings happen to be compatible with
observational cosmology if they form at the GUT scale. It was
shown however by Witten~\cite{witten} that, depending on the
explicit realization of the symmetry breaking scheme as well as
on the various particle couplings, a current could build along
the strings, thereby effectively turning them into
superconducting wires. Such wires were originally considered in
the case the current couples to the electromagnetic field so they
may be responsible for a variety of new effects, including an
explosive scenario for large scale structure formation for which
an enormous energy release was realized in the form of an
expanding shell of non propagating photons in the surrounding
plasma~\cite{otw}.

The cosmology of strings has been the subject of intense work in
the last twenty years or so~\cite{strings}, mainly based on
ordinary strings, global or local, aiming at deriving the large
scale structure properties stemming from their distribution as
well as their imprint in the microwave background~\cite{CMBR}. It
was even shown~\cite{BPRS} that the most recent data~\cite{BOOM}
might support a non negligible contribution of such defects. As
such a result requires ordinary strings, it turns out to be of
uttermost importance to understand the influence of currents in
the cosmological context.

Indeed, it can be argued that currents might drastically modify
the cosmological evolution of a string network: The most clearly
defined consequence of the existence of a current flowing along a
vortex is the breaking of the boost invariance, since the current
itself defines a privileged frame. In other words, the energy per
unit length $U$ and the tension $T$ become two different numbers,
contrary to the ordinary (Goto-Nambu~\cite{gn}) case. As a
result, string loops become endowed with the capability of
rotation (the latter being meaningless for $U=T$), and the induced
centrifugal force permits equilibrium configurations, called
vortons~\cite{davisRL}. They would very rapidly reach a regime
where they would scale as ordinary non relativistic matter, until
they come to completely dominate the
Universe~\cite{brandenberger}.

In the original Witten model~\cite{witten}, currents could form by
means of two different mechanisms. Scalar fields, directly
coupled with the string forming Higgs field, could feel a
localized potential into which they could accumulate in the form
of bound states, while fermions could be trapped along the
string, propagating at the speed of light, as zero energy
solutions of the two dimensional Dirac equation around the
vortex. Other models were proposed where fermions could also
propagate in the string core at lower velocities in the form of
massive modes~\cite{massF}, or (possibly charged) vector fields
could also condense~\cite{vecmass}. All these models have
essentially made clear that the existence of currents in string
is much more than a mere possibility but rather an almost
unavoidable fact in realistic particle physics theories.

For scalar as well as vector carriers, the task of understanding
the microphysics is made simple thanks to their bosonic nature:
all the trapped particles go into the same lowest accessible
energy state and the field can be treated
classically~\cite{bps,neutral}. Even the surrounding
electromagnetic~\cite{enon0} and
gravitational~\cite{garrigapeter,peterpuy} fields generated by the
current can be treated this way and the backreaction can be
included easily~\cite{nospring}.

Meanwhile, a general formalism was set up by Carter~\cite{formal}
to describe current-carrying string dynamics. The formalism is
based on a single so-called state parameter, $w$ say, of which the
energy per unit length, the tension and the current itself are
functions. Such a formalism relied heavily on the fact that for a
bosonic carrier, the relevant quantity whose variation along the
string leads to a current is its phase, and the state parameter
is essentially identifiable to this phase gradient. Various
equations of state relating the tension to the energy per unit
length were then derived~\cite{cp1}, based on numerical results
and the existence of a phase frequency threshold~\cite{neutral}.
It even includes the special case of a chiral current~\cite{cp2},
although the latter originates in principle only for a purely
fermionic current.  Therefore, it was until now implicitly assumed
that such a formalism would be sufficient to describe whatever
current-carrying string configuration. It will be shown in this
paper that this is in fact not true and an extended version,
including more than one parameter, is needed~\cite{prep}.

The state parameter formalism, apart from being irrelevant
for fermionic current-carrying strings, can only provide a
purely classical description of their dynamics. This is
unfortunate since the most relevant prediction of superconducting
cosmic string models in cosmology is the existence of the vorton
states discussed above. These equilibrium configurations of
rotating loops are not necessarily stable, and in fact, this is
perhaps the most important question to be answered on this topic.
Indeed, any theory leading to the the existence of absolutely stable
vortons predicts a cosmological catastrophe~\cite{brandenberger}
and must be ruled out.
One may therefore end up with a very stringent constraint
on particle physics extension of the standard model of electroweak
and strong interactions. To decide clearly on this point requires
to investigate both the classical and the quantum stability of
vortons.

Classical stability has already been established in the case of
bosonic carriers~\cite{martinpeter} for whatever equation of
state~\cite{cartermartin} on the basis of the one parameter
formalism. Yet it will also have to be addressed in the more
general context that will be discussed below. In the meantime
it was believed that a quantum treatment was necessary in order
to decide on the quantum stability: as one wants to compare the
characteristic life-time of a vorton with the age of the
Universe, quantum effects can turn out to be relevant; hence the
following work in which the simplest of all fermionic Witten
models is detailed~\cite{witten} that can give rise to both
spacelike as well as timelike charge currents, generalizing
the usual point of view~\cite{davisC}.

Let us sketch the lines along which this work is made.

A two-dimensional quantization of the spinor fields involved
along a string is performed. Owing to anti-particle exitation
states, one can derive the conditions under which the current is
of arbitrary kind. Moreover, an equation of state giving the
energy per unit length and the tension is obtained that involves
four different state parameters which are found to be the number
densities of fermions, although three of them only happen to be
independent.

In section \ref{modele}, the model is presented and motivated, and
the equations of motion are derived. Then in section
\ref{modezero}, we obtain plane wave solutions along the string
by separating transverse and longitudinal dependencies of spinor
fields in the vortex. The zero mode transverse solutions are then
constraints to be normalizable in order to represent well defined
wave functions. The quantization restricted to massless
longitudinal modes is performed in section \ref{quantization}.
As a result, the classical conserved currents obtained from Noether
theorem, like energy-momentum tensor and fermionic currents, are
expressed in their quantum form.
All these quantum operators end up being
functions of the fermionic occupation numbers only. In the last
section (section \ref{etat}), the classical expressions for the
energy per unit length and the tension are derived and discussed
from computation of quantum observable values of the stress
tensor operator in the classical limit. Contrary to the bosonic
current-carrier case where there is only one state
parameter~\cite{neutral}, the classical limit of the model
involves four state parameters in order to fully determine the
energy per unit length and the tension. The cosmological
consequences of this new analysis are briefly discussed in the
concluding section.

\section{Equations of motion}
\label{modele}

We are going to be interested in the purely dynamical effects a
fermionic current flowing along a cosmic string may have. The
model we will be dealing with here is a simplified version of
that proposed by Witten~\cite{witten} which involves two kinds of
fermions, in the neutral limit. This limit, for which the
coupling between fermions and electromagnetic-like external fields
is made to vanish, permits an easy recognition of the dynamical
effects of the existence of an internal structure as in
Ref.~\cite{neutral}.

\subsection{Particle content}

The model we shall consider involves a complex scalar Higgs field,
$\Phi$ say, with conserved charge $qc_{\Phi}$ under a local $U(1)$
symmetry, together with the associated gauge vector field
$B^{\mu}$. In this simple Abelian Higgs model~\cite{higgs},
vortices can form after spontaneous breaking of the $U(1)$
symmetry. The minimal anomaly free model~\cite{witten} with spinor
fields requires two Dirac fermions denoted $\Psi$ and $\Chi$, with
opposite electromagnetic-like charges, getting their masses
from chiral coupling with the Higgs field and its complex
conjugate. They also have conserved gauge charges from invariance
under the broken $U(1)$ symmetry, $q\cpsir$, $q\cpsil$,
$q\cchir$, and $q\cchil$, for the right- and left-handed parts of
the two fermions respectively. The Lagrangian of the model
therefore reads
\begin{equation}
{\mathcal{L}}={\mathcal{L}}_h+{\mathcal{L}}_g+{\mathcal{L}}_\psi+
{\mathcal{L}}_\chi
\end{equation}
with ${\mathcal{L}}_h$, ${\mathcal{L}}_g$ and
${\mathcal{L}}_\psi$, ${\mathcal{L}}_\chi$, respectively the
Lagrangian in the Higgs, gauge, and fermionic sectors. In terms
of the underlying fields, they are
\begin{eqnarray}
{\mathcal{L}}_h & = & \frac{1}{2} (D_{\mu}\Phi)^{\dag}(D^{\mu}\Phi)
- V(\Phi),
\\
{\mathcal{L}}_g & = & -\frac{1}{4} H_{\mu \nu} H^{\mu \nu},
\\
\label{psilagrangian}
{\mathcal{L}}_\psi & = & \frac{i}{2} \left[\overline{\Psi}
\gamma^{\mu} D_{\mu} \Psi - (\overline{D_{\mu}\Psi}) \gamma^{\mu}
\Psi \right] -g \overline{\Psi} \frac{1+\gamma_5}{2} \Psi \Phi
-g\overline{\Psi} \frac{1-\gamma_5}{2} \Psi \Phi^\ast,
\\
\label{chilagrangian}
{\mathcal{L}}_\chi & = & \frac{i}{2} \left[\overline{{\Chi}}
\gamma^{\mu} D_{\mu} {\Chi} - (\overline{D_{\mu}{\Chi}})
\gamma^{\mu} {\Chi} \right] -g \overline{{\Chi}}
\frac{1+\gamma_5}{2} {\Chi} \Phi^{\ast} - g \overline{{\Chi}}
\frac{1-\gamma_5}{2} {\Chi} \Phi,
\end{eqnarray}
where we have used the notation
\begin{eqnarray}
D_{\mu}\Phi & = & (\nabla_{\mu} + i q c_{\phi}B_\mu) \Phi,
\\
D_{\mu}\Psi & = & (\nabla_{\mu} + i q \frac{\cpsir+\cpsil}{2}B_\mu
+i q \frac{\cpsir-\cpsil}{2} \gamma_5 B_\mu) \Psi,
\\
D_{\mu}{\Chi} & = & (\nabla_{\mu} + i q \frac{\cchir+
\cchil}{2}B_\mu +i q \frac{\cchir-\cchil}{2} \gamma_5 B_\mu)
{\Chi},
\\
H_{\mu \nu} & = & \nabla_\mu B_\nu - \nabla_\nu B_\mu,
\\
V(\Phi) & = & \frac{\lambda}{8} (|\Phi|^2 - \eta^2)^2.
\end{eqnarray}
The equivalence with the Witten model~\cite{witten} appears through a
separation into left- and right-handed spinors. Let us define
$\Psi_R$ and $\Psi_L$, respectively the right- and left-handed parts
of the Dirac spinor field $\Psi$ (and the same for ${\Chi}$),
eigenvectors of $\gamma_5$,
\begin{eqnarray}
\Psi_R  =  \frac{1+\gamma_5}{2} \Psi,
& \quad \hbox{\textrm{and}} \quad &
\Psi_L  =  \frac{1-\gamma_5}{2} \Psi.
\end{eqnarray}
The Lagrangian for the spinor field $\Psi$ now reads
\begin{eqnarray}
{\mathcal{L}}_\psi & = & \frac{i}{2} \left[\overline{\Psi}_R
\gamma^{\mu} D_{\mu} \Psi_R - (\overline{D_{\mu}\Psi}_R) \gamma^{\mu}
\Psi_R \right] + \frac{i}{2} \left[\overline{\Psi}_L \gamma^{\mu} D_{\mu}
\Psi_L - (\overline{D_{\mu}\Psi}_L) \gamma^{\mu} \Psi_L \right]
-g \overline{\Psi}_L \Psi_R \Phi -g\overline{\Psi}_R \Psi_L \Phi^\ast,
\end{eqnarray}
with the associated covariant derivatives
\begin{equation}
D_{\mu}\Psi_{R(L)} = (\nabla_{\mu} + i qc_{\psi_{R(L)}}B_\mu) \Psi_{R(L)}.
\end{equation}
It is clear with the Lagrangian expressed in this way that the
invariance of the action under $U(1)$ transformations requires
\begin{equation}
\cpsil-\cpsir=c_\phi=\cchir-\cchil.
\end{equation}

\subsection{Equations of motion}

As we wish to deal with a cosmic string, the Higgs and gauge
fields can be set as a vortex-like Nielsen--Olesen solution and
they can be written in cylindrical coordinates as
follows~\cite{NO}
\begin{equation}
\begin{array}{ccc}
\Phi  =  \varphi(r) \ue^{i \alpha(\theta)},
& \quad &
B_\mu  =  B(r) \delta_{\mu \theta}.
\end{array}
\end{equation}
In order for the Higgs field to be well defined by rotation around
the string, its phase has to be proportional to the orthoradial
coordinate, $\alpha(\theta)=n \theta$, where the integer $n$ is the
winding number. The new fields $\varphi$ and $\alpha$ are now real
scalar fields and are solutions of the equations of motion
\begin{eqnarray}
\label{higgsmvt}
\nabla_\mu \nabla^\mu \varphi & = & \varphi Q_\mu Q^\mu -
\frac{\ud V(\varphi)}{\ud \varphi}
- \overline{\Psi} \frac {\partial m_\psi}{\partial \varphi} \Psi
- \overline{{\Chi}} \frac {\partial m_\chi}{\partial \varphi}
{\Chi},
\\
\nabla_\mu \left(\varphi^2 Q^\mu \right) & = &
-\overline{\Psi} \frac {\partial m_\psi}{\partial \alpha} \Psi
- \overline{{\Chi}} \frac {\partial m_\chi}{\partial \alpha}
{\Chi},
\end{eqnarray}
where
\begin{eqnarray}
Q_\mu & = & \nabla_\mu \alpha + q c_\phi B_\mu,
\\
m_\psi & = & g \varphi \cos{\alpha} + i g \varphi \gamma_5 \sin{\alpha},
\\
m_\chi & = & g \varphi \cos{\alpha} - i g \varphi \gamma_5 \sin{\alpha}.
\end{eqnarray}
In the same way, the equations of motion for the gauge and spinor
fields are
\begin{eqnarray}
\label{gaugemvt}
\nabla_\mu H^{\mu \nu} & = & j^\nu_\psi + j^\nu_\chi - q c_\phi \varphi^2
Q^\nu,
\\
\label{psimvt}
i \gamma^\mu \nabla_\mu \Psi & = & \frac{\partial j^\mu_\psi}{\partial
\overline{\Psi}} B_\mu + m_\psi \Psi,
\\
\label{psibarmvt}
i \left(\nabla_\mu \overline{\Psi}\right) \gamma^\mu & = &
- \frac{\partial j^\mu_\psi}{\partial \Psi} B_\mu - \overline{\Psi} m_\psi,
\\
\label{chimvt}
i \gamma^\mu \nabla_\mu {\Chi} & = & \frac{\partial j^\mu_\chi}
{\partial \overline{{\Chi}}} B_\mu + m_\chi {\Chi},
\\
\label{chibarmvt}
i \left(\nabla_\mu \overline{{\Chi}} \right) \gamma^\mu & = &
- \frac{\partial j^\mu_\chi}{\partial {\Chi}} B_\mu
-\overline{{\Chi}} m_\chi.
\end{eqnarray}
The fermionic currents $j^\mu_{\psi(\chi)}$ have axial and
vectorial components due to the different coupling between left-
and right-handed spinors to the gauge field. The two kinds of
current are required to respect gauge invariance of the
Lagrangian. In terms of spinor fields, they read
\begin{equation}
j^\mu_{\psi(\chi)}=j^\mu_{\psi_V(\chi_V)}+j^\mu_{\psi_A(\chi_A)},
\end{equation}
with
\begin{equation}
\label{currents}
\begin{array}{ccccccclll}
j^\mu_{\psi_V} & = & q \displaystyle{\frac{\cpsir+ \cpsil}{2}}
\overline{\Psi} \gamma^\mu \Psi, & \quad & j^\mu_{\chi_V} & = & q
\displaystyle{\frac{\cchir+ \cchil}{2}} \overline{{\Chi}} \gamma^\mu {\Chi},
\\ \\
j^\mu_{\psi_A} & = & q \displaystyle{\frac{\cpsir- \cpsil}{2}}
\overline{\Psi} \gamma^\mu \gamma_5 \Psi, & \quad &
j^\mu_{\chi_A} & = & q \displaystyle{\frac{\cchir- \cchil}{2}}
\overline{{\Chi}} \gamma^\mu \gamma_5 {\Chi}.
\end{array}
\end{equation}

\section{Transverse solutions as Zero Modes}
\label{modezero}

\subsection{Plane wave solutions}

The study of the fermionic fields trapped along the string can be
performed by separating the longitudinal and transverse solutions
of the equations of motion. The plane wave solutions are
therefore expressed in the generic form
\begin{equation}
\label{planeansatz}
\begin{array}{lll}
\Psi_p^{(\pm)}  =  \ue^{\pm i(\omega t-kz)}
\left(\begin{array}{l}
\xi_1(r) \ue^{-im_1 \theta} \\
\xi_2(r) \ue^{-im_2 \theta} \\
\xi_3(r) \ue^{-im_3 \theta} \\
\xi_4(r) \ue^{-im_4 \theta}
\end{array} \right),
& \quad &
{\Chi}_p^{(\pm)}  =  \ue^{\pm i(\omega t-kz)}
\left(\begin{array}{l}
\zeta_1(r) \ue^{-il_1 \theta} \\
\zeta_2(r) \ue^{-il_2 \theta} \\
\zeta_3(r) \ue^{-il_3 \theta} \\
\zeta_4(r) \ue^{-il_4 \theta}
\end{array} \right).
\end{array}
\end{equation}
Inside the vortex, the numbers $m_i$ and $l_i$ have to be
integers in order to produce well defined spinors by rotation
around the string. In the following, the Dirac spinors will be
expressed in the chiral representation, and the metric is assumed
to have the signature $(+,-,-,-)$. Plugging the expression
(\ref{planeansatz}) into the equations of motion (\ref{psimvt})
and (\ref{chimvt}) yields the differential system
\begin{equation}
\label{syst}
\left\{
\begin{array}{lll}
\vspace{4pt}
\displaystyle
\left(\frac{\ud\xi_1}{\ud r} + \frac{1}{r}\left[-q \cpsir B(r) + m_1\right]
\xi_1(r) \right) \ue^{-i(m_1-1)\theta} - i g \varphi(r) \ue^{-i(m_4+n)\theta}
\xi_4(r) & = & \mp i (k+\omega) \xi_2(r) \ue^{-im_2\theta}, \\
\vspace{4pt}
\displaystyle
\left(\frac{\ud\xi_2}{\ud r} + \frac{1}{r}\left[q \cpsir B(r) - m_2\right]
\xi_2(r) \right) \ue^{-i(m_2+1)\theta} - i g \varphi(r) \ue^{-i(m_3+n)\theta}
\xi_3(r) & = & \pm i (k-\omega) \xi_1(r) \ue^{-im_1\theta}, \\
\vspace{4pt}
\displaystyle
\left(\frac{\ud\xi_3}{\ud r} + \frac{1}{r}\left[-q \cpsil B(r) + m_3\right]
\xi_3(r) \right) \ue^{-i(m_3-1)\theta} + i g \varphi(r) \ue^{-i(m_2-n)\theta}
\xi_2(r) & = & \mp i (k-\omega) \xi_4(r) \ue^{-im_4\theta}, \\
\displaystyle
\left(\frac{\ud\xi_4}{\ud r} + \frac{1}{r}\left[q \cpsil B(r) - m_4\right]
\xi_4(r) \right) \ue^{-i(m_4+1)\theta} + i g \varphi(r) \ue^{-i(m_1-n)\theta}
\xi_1(r) & = & \pm i (k+\omega) \xi_3(r) \ue^{-im_3\theta}.
\end{array}
\right.
\end{equation}
Similar equations are obtained for the field ${\Chi}$ with the
following transformations, $\xi \rightarrow \zeta$, $c_{\psi_{R(L)}}
\rightarrow c_{\chi_{R(L)}}$, and $n \rightarrow -n$, because of its
coupling to the anti-vortex instead of the vortex. Note that if,
instead of the vectorial phases ansatz (\ref{planeansatz}), we had chosen
a matricial phases ansatz in the form
\begin{equation}
\Psi_p^{(\pm)}  =  \ue^{\pm i(\omega t-kz)}
\left(\begin{array}{llll}
\xi_{11}(r) \ue^{-im_{11} \theta}+ \ldots + \xi_{14}(r) \ue^{-im_{14} \theta} \\
\xi_{21}(r) \ue^{-im_{21} \theta}+ \ldots + \xi_{24}(r) \ue^{-im_{24} \theta} \\
\xi_{31}(r) \ue^{-im_{31} \theta}+ \ldots + \xi_{34}(r) \ue^{-im_{34} \theta} \\
\xi_{41}(r) \ue^{-im_{41} \theta}+ \ldots + \xi_{44}(r) \ue^{-im_{44} \theta}
\end{array} \right),
\end{equation}
we would have found, from the requirement of having at most four independent
phases in the equations of motion, that the matrix $m_{ij}$ has to verify
for all $i$, $m_{ij}=m_{ik}$, for all $(j,k)$. Consequently, the vectorial
ansatz (\ref{planeansatz}) is the most general for solutions with separated
variables.

\subsection{Transverse solutions}

{}From the differential system (\ref{syst}), it is obvious that
the four phases $m_i$ cannot be independent parameters if the
spinors fields are not identically zero. It is also impossible to
find three independent phase parameters since each equation
involves precisely three different angular dependencies. The only
allowed angular separation requires two degrees of freedom in
$\theta$, and the only relevant relation for trapped modes in the
string reads from Eq.~(\ref{syst})
\begin{eqnarray}
\label{angseparation}
m_1-1=m_4+n,
& \quad &
m_2+1=m_3+n.
\end{eqnarray}

\subsubsection{Zero modes}

Introducing the two integer parameters $p=m_1$ and
$q=m_3$, and using equation (\ref{angseparation}), for $p\neq q+n$,
the system (\ref{syst}) reduces to the set
\begin{equation}
\label{zeromodes}
\left\{
\begin{array}{l}
\vspace{4pt}
\displaystyle
\frac{\ud \xi_2}{\ud r} + \frac{1}{r}\left[q \cpsir B(r) - (q+n-1)\right]
\xi_2(r) - i g \varphi(r) \xi_3(r)  = 0, \\
\vspace{4pt}
\displaystyle
\frac{\ud \xi_3}{\ud r} + \frac{1}{r}\left[-q \cpsil B(r) + q\right]
\xi_3(r) + i g \varphi(r) \xi_2(r) = 0, \\
\vspace{4pt}
\displaystyle
(k+\omega) \xi_2(r) = 0, \\
\displaystyle
(k+\omega) \xi_3(r) = 0,
\end{array}
\right.
\end{equation}
\begin{equation}
\label{antizeromodes}
\left\{
\begin{array}{l}
\vspace{4pt}
\displaystyle
\frac{\ud \xi_1}{\ud r} + \frac{1}{r}\left[-q \cpsir B(r) + p\right]
\xi_1(r) - i g \varphi(r)\xi_4(r) = 0, \\
\vspace{4pt}
\displaystyle
\frac{\ud \xi_4}{\ud r} + \frac{1}{r}\left[q \cpsil B(r) - (p-n-1)\right]
\xi_4(r) + i g \varphi(r) \xi_1(r) = 0, \\
\vspace{4pt}
\displaystyle
(k-\omega) \xi_1(r) = 0, \\
\displaystyle
(k-\omega) \xi_4(r) = 0.
\end{array}
\right.
\end{equation}
There are two kinds of solutions which propagate along the two directions
of the string at the speed of light and which were originally found by Witten
\cite{witten}:
Either $k=\omega$ and $\xi_2 =\xi_3=0$, or $k=-\omega$ and $\xi_1=\xi_4=0$.
These zero modes must also be normalizable in the transverse plane of the
string in order to be acceptable as wave functions.

\subsubsection{Index theorem}

In the two cases $k=\omega$ and $k=-\omega$, we will call the
corresponding zero modes, $X_\psi(r,\theta)$ and
$Y_\psi(r,\theta)$, the solutions of the systems
(\ref{zeromodes}) and (\ref{antizeromodes}), respectively, i.e.,
\begin{equation}
\begin{array}{lll}
X_\psi = \left(
\begin{array}{l}
0 \\
\xi_2(r) \ue^{-i (q+n-1) \theta} \\
\xi_3(r) \ue^{-i q \theta} \\
0
\end{array}
\right),
& \quad &
Y_\psi = \left(
\begin{array}{l}
\xi_1(r) \ue^{-i p \theta} \\
0 \\
0 \\
\xi_4(r) \ue^{-i (p-n-1) \theta}
\end{array}
\right).
\end{array}
\end{equation}
These have to be normalizable in the sense that $\int{|X_\psi|^2
r\, \ud r \,\ud\theta}$ and $\int{|Y_\psi|^2 r\,\ud
r\,\ud\theta}$ must be finite. Thanks to the regularity of the
vortex background, the divergences in these integrals can only
arise close to the string core or asymptotically far away from it.
As a result, it is sufficient to study asymptotic behaviors of
the solutions to decide on their normalizability~\cite{jackiwrossi}. \\
Let us focus on the zero mode $X_\psi$ solution of
Eq.~(\ref{zeromodes}), keeping in mind that $Y_\psi$ can be dealt
with in the same way. The asymptotic behaviors at infinity are
easily found as solutions of the limit at infinity of the
differential system (\ref{zeromodes}). Note that the
identification between the equivalent solutions and the solutions
of the equivalent system is only allowed by the absence of
singular point for the system at infinity, and it will not be so
near the string since $r=0$ is a singular point, so that the
Cauchy theorem does no longer apply. From Eq.~(\ref{zeromodes}), the
eigensolutions of the equivalent system at infinity are in the form
$\exp{(\pm g \eta r)}$, and thus, there is only one normalizable
solution at infinity.

Near the string, i.e., where $r \rightarrow 0$, the system is no
longer well defined, the origin being a singular point.
Approximate solutions can however be found by looking at the
leading term of a power-law expansion of both system and
functions, as originally suggested by Jackiw and Rossi
\cite{jackiwrossi}. Because the Cauchy theorem does no longer
apply, many singular solutions might be found at the origin, and
among them, the two generic ones which match with the two
exponentials at infinity. Near the origin, the Higgs and gauge
fields are known to behave like~\cite{NO}
\begin{equation}
\label{equizero}
\begin{array}{lll}
\varphi(r) \sim \varphi_o r^{|n|}, & \quad & B(r) \propto r^2,
\end{array}
\end{equation}
so that the leading contribution near the string of the zero modes
can be found as
\begin{eqnarray}
\xi_2(r) & \sim & a_2 r^{\alpha_2}, \\
\xi_3(r) & \sim & a_3 r^{\alpha_3},
\end{eqnarray}
with $a_i$ and $\alpha_i$ real parameters to be determined.
The values of the exponents $\alpha_i$ are therefore given by the
leading order terms in system (\ref{zeromodes}), and one obtains
three solutions, the first one of which being singular,
\begin{equation}
\begin{array}{lllll}
\left(
\begin{array}{l}
\xi_2 \\
\xi_3
\end{array}
\right)_s
=
\left(
\begin{array}{l}
a_2 r^{q+n-1} \\
a_3 r^{-q}
\end{array}
\right),
& \quad & \textrm{provided} & \quad &
\displaystyle
-\frac{|n|+n}{2} < q <1+ \frac{|n|-n}{2}.
\end{array}
\end{equation}
The two other solutions are the generic ones which have to match with
the solutions at infinity
\begin{equation}
\begin{array}{lllllllll}
\left(
\begin{array}{l}
\xi_2 \\
\xi_3
\end{array}
\right)_{g1}
& = &
\left(
\begin{array}{l}
a_2(a_3) r^{-q+|n|+1} \\
a_3 r^{-q}
\end{array}
\right),
& \quad & \textrm{and} & \quad &
\left(
\begin{array}{l}
\xi_2 \\
\xi_3
\end{array}
\right)_{g2}
& = &
\left(
\begin{array}{l}
a_2(a_3) r^{q+n-1} \\
a_3 r^{q+|n|+n}
\end{array}
\right),
\end{array}
\end{equation}
where the relationships between the parameters $a_i$ have not been written,
since they are clearly obtained from Eq.~(\ref{zeromodes}).
Normalizability near the origin requires that the integrals $\int{|\xi_2|^2\,
r\, \ud r}$ and $\int{|\xi_3|^2 \,r \,\ud r}$ converge, and this yields
\begin{equation}
\label{criterionzeromodes}
-n < q < 1.
\end{equation}
Analogous considerations for the system (\ref{antizeromodes}) show the
convergence criterion in this case to be
\begin{equation}
\label{criterionantizeromodes}
n < p < 1.
\end{equation}
The number of sets of parameters $p$ and $q$ satisfying the
previous inequalities is precisely the number of well defined
zero modes, respectively $Y_\psi$ and $X_\psi$, which are also
normalizable. In order to match with the single well behaved
solution at infinity, the two independent solutions near the
string have to be integrable. Therefore, for a vortex solution
with a positive winding number $n$, there are only $n$
normalizable zero modes, which are the $X_\psi$ ones. Similarly
in the case of an anti-vortex with negative winding number
$-|n|$, one finds  also $|n|$ zero modes $Y_\psi$. This is the
index theorem found by Jackiw and Rossi~\cite{jackiwrossi}.
Recall that the model involves two kinds of fermions, and all the
previous considerations apply as well for the field ${\Chi}$ with
the simple transformation $n \rightarrow -n$. Therefore, the
normalizable zero modes are swapped compared to those of the
field $\Psi$.

Finally, for a vortex with positive winding number
$n$, there are always $n$ massless plane wave solutions for both
spinor fields, which read
\begin{equation}
\label{psisolutions}
\Psi_p^{(\pm)} = \ue^{\pm i k(t+z)} \left(
\begin{array}{llll}
0 \\
\xi_2(r) \ue^{-i(q+n-1) \theta} \\
\xi_3(r) \ue^{-iq \theta}\\
0
\end{array}
\right)
=  \ue^{\pm i k(t+z)} X_\psi(r,\theta),
\end{equation}
\begin{equation}
\label{chisolutions}
{\Chi}_p^{(\pm)} = \ue^{\pm ik(t-z)} \left(
\begin{array}{llll}
\zeta_1(r) \ue^{-ip \theta} \\
0 \\
0 \\
\zeta_4(r) \ue^{-i(p+n-1) \theta}
\end{array}
\right)
=  \ue^{\pm ik(t-z)} Y_\chi(r,\theta),
\end{equation}
with now $q=m_3$ and $p=l_1$ which satisfy
\begin{equation}
\begin{array}{ccccc}
-n < q < 1, & \quad & \textrm{and} & \quad & -n < p < 1.
\end{array}
\end{equation}
Note that they are eigenvectors of the $\gamma^0 \gamma^3$ operator,
and they basically verify
\begin{equation}
\label{currentszeromodes}
\begin{array}{llllllllllllllllll}
\overline{X_\psi} \gamma^0 X_\psi & = & |X_\psi|^2,
& \quad &
\overline{X_\psi}\gamma_5 \gamma^0 X_\psi & = & -|\xi_2|^2+|\xi_3|^2,
& \quad &
\overline{X_\psi}\gamma^3 X_\psi & = & -|X_\psi|^2,
\\
\\
\overline{X_\psi} \gamma^{1(2)} X_\psi & = & 0,
& \quad &
\overline{X_\psi} \gamma_5 \gamma^3 X_\psi & = & |\xi_2|^2-|\xi_3|^2,
& \quad &
\overline{X_\psi} \gamma_5 \gamma^{1(2)} X_\psi & = & 0,
\\
\\
\overline{Y_\psi} \gamma^0 Y_\psi & = & |Y_\psi|^2,
& \quad &
\overline{Y_\psi} \gamma_5 \gamma^0 Y_\psi & = & -|\xi_1|^2+|\xi_4|^2,
& \quad &
\overline{Y_\psi} \gamma^3 Y_\psi & = & |Y_\psi|^2,
\\
\\
\overline{Y_\psi} \gamma^{1(2)} Y_\psi & = & 0,
& \quad &
\overline{Y_\psi} \gamma_5 \gamma^3 Y_\psi & = & -|\xi_1|^2+|\xi_4|^2,
& \quad &
\overline{Y_\psi} \gamma_5 \gamma^{1(2)} Y_\psi & = & 0.
\end{array}
\end{equation}
The ${\Chi}$ zero modes $X_\chi$ and $Y_\chi$ verify the same
relationships with $\xi_i$ replaced by $\zeta_i$.

\subsubsection{Massive modes}

The case $m_1=m_3+n$ allows four-dimensional solutions of the
system (\ref{syst}). In Particular, these solutions do no longer
require $\omega=\pm k$ and therefore represent massive modes. As
before, the interesting behaviors of these modes are found by
studying the solutions of the equivalent system asymptotically
and by looking for the leading term of a power-law expansion of
both system and solution near the string core.

At infinity, the system is well defined and there are two twice
degenerate eigensolutions $\exp{(\pm \Omega r)}$ out of which two
are normalizable, with
\begin{equation}
\Omega = \sqrt{g^2 \eta^2-(\omega^2-k^2)}.
\end{equation}
Near the origin, at $r=0$, the system is singular, and because of
its four dimensions there are much more singular solutions than
the previous two dimensional case, and among them the four
generic ones which match with the four ones at infinity. The
leading term in asymptotic expansion can be written in a standard
way
\begin{equation}
\xi_i(r) \sim a_i r^{\alpha_i}.
\end{equation}
Plugging these expressions in the system (\ref{syst}) with
Eq.~(\ref{equizero}), and keeping only leading terms at $r=0$ gives,
after some algebra, the four generic solutions
\begin{equation}
\label{onephasemodes}
\begin{array}{lllllll}
\left(
\begin{array}{l}
\xi_1 \\
\xi_2 \\
\xi_3 \\
\xi_4
\end{array}
\right)_{g1}
 & = &
\left(
\begin{array}{l}
a_1 r^{-m} \\
a_2(a_1) r^{-m+1} \\
a_3(a_1) r^{-m+|n|+2} \\
a_4(a_1) r^{-m+|n|+1}
\end{array}
\right),
& \quad &
\left(
\begin{array}{l}
\xi_1 \\
\xi_2 \\
\xi_3 \\
\xi_4
\end{array}
\right)_{g2}
& = &
\left(
\begin{array}{l}
a_1 r^{m+|n|-n} \\
a_2(a_1) r^{m+|n|-n+1} \\
a_3(a_1) r^{m-n} \\
a_4(a_1) r^{m-n-1}
\end{array}
\right),
\\ \\
\left(
\begin{array}{l}
\xi_1 \\
\xi_2 \\
\xi_3 \\
\xi_4
\end{array}
\right)_{g3}
 & = &
\left(
\begin{array}{l}
a_1 r^{m} \\
a_2(a_1) r^{m-1} \\
a_3(a_1) r^{m+|n|} \\
a_4(a_1) r^{m+|n|+1}
\end{array}
\right),
& \quad &
\left(
\begin{array}{l}
\xi_1 \\
\xi_2 \\
\xi_3 \\
\xi_4
\end{array}
\right)_{g4}
& = &
\left(
\begin{array}{l}
a_1 r^{-m+|n|+n+2} \\
a_2(a_1) r^{-m+|n|+n+1} \\
a_3(a_1) r^{-m+n} \\
a_4(a_1) r^{-m+n+1}
\end{array}
\right),
\end{array}
\end{equation}
with $m=m_1$, and where the relationships between the coefficients
$a_i$ have not been written as they are essentially given by a
linear system in $a_i$ given by Eq.~(\ref{syst}).
The solutions (\ref{onephasemodes}) will be normalizable near the
string if, for all $i$, $\int{|\xi_i|^2\, r\, \ud r}$ is finite.
Moreover there will be, at least, always one massive bound state
if there are at least three normalizable eigensolutions to match
with the well-defined ones at infinity. This is only allowed if
the parameter $m$ verifies simultaneously three of the following
conditions
\begin{equation}
\sup{(0,n)} < m < \inf{(1,1+n)}.
\end{equation}
Because $m$ is necessary an integer, this condition cannot be
achieved. This criterion, originally derived and used by Jackiw
and Rossi in order to enumerate the number of zero modes in a
vortex-fermion system~\cite{jackiwrossi}, is only sufficient and
thus, normalizable massive bound states may exist, but are model
dependent since it is necessary that a particular combination of
the two normalizable eigenmodes near the string core match
exactly with a particular combination of the two well-defined
ones at infinity.

Such massive bound states depend therefore of the particular
values of the model parameters. Recently, it was shown
numerically~\cite{davisC} that the Abelian Higgs model with one
Weyl fermion admits always at least two massive bound states, as
a result, the present toy model also may have such states.
However, in order to simplify the quantization, we will only
consider the generic zero modes, and consequently, the following
results will be relevant for cosmic string only when the
occupancy of the massive bound states can be neglected compared
to the occupancy of the zero mode states. Such physical
situations are likely to occur far below the energy scale where
the string was formed, since the massive states are generally
expected to decay much more rapidly than the massless ones
\cite{davisC}.

The generic massless normalizable transverse solutions of the
fermionic equations of motion in the string with winding number
$n$ are the $n$ zero modes. For the spinor field $\Psi$ coupled
with the vortex, we find that the particles and the
anti-particles can only propagate at the speed of light in one
direction, ``$-z$'' say direction along the string, whereas the
spinor field ${\Chi}$ propagates in the opposite, ``$+z$''
direction. The existence of such plane waves allows us to
quantize the spinor fields along the string. The zero modes
themselves will therefore be transverse wave functions giving the
probability density for finding a trapped mode at a chosen
distance from the string core.

\section{Fock space along the string}
\label{quantization}

The spinor fields can be expanded on the basis of the plane wave
solutions computed above. A canonical quantization can then be
performed along the $z$-axis which provides analytical
expressions for these fields in two dimensions once the
transverse degrees of freedom have been integrated over. It is
therefore possible to compute the current operators as well as
their observable values given by their averages in a particular
Fock state. In the following we shall take a vortex with a unit
winding number $n=1$ and the subscript of the zero modes $X_\psi$
and $Y_\chi$ will be forgotten since there is not possible
confusion.

\subsection{Canonical quantization}

We shall first be looking for a physical expansion of the spinor
fields in plane waves, in the sense that creation and
annihilation operators are well defined. The Hamiltonian is then
calculable as a function of these and will be required to be
positive to yield a reasonable theory.

\subsubsection{Quantum fields}

As shown above, the spinor fields $\Psi$ and ${\Chi}$ propagate in
only one direction, therefore in expressions (\ref{psisolutions})
and (\ref{chisolutions}) the momentum $k$ can be chosen positive
definite. Let us, once again, focus on the spinor field $\Psi$.
The natural way to expand it in plane waves of positive and
negative energies is
\begin{equation}
\label{psiexpansion}
\Psi=\int_{0}^{\infty}{\frac{\ud k}{2\pi2k}\left[b^{\dag}(-k) \ue^{ik(t+z)} +
\underline{b}(-k) \ue^{-ik(t+z)} \right] X}.
\end{equation}
The Fourier transform of $\Psi$ on positive and negative energies
has been written with similar notation $b^\dag$ and
$\underline{b}$ unlike in the free spinor case. Indeed, note
that, in the string, the zero modes are the same for both
positive and negative energy waves, so that the only way to
distinguish particles from anti-particles is in the sign of the
energy. The integration measure $\ud k/(2\pi2k)$ is the usual
Lorentz invariant measure in two dimensions. Note that $k$ is
chosen always positive in order to represent physical energy and
momentum actually carried by the field along the string; hence the
negative sign in $b^\dag(-k)$ and $\underline{b}(-k)$, which is a
reminder that the spinor field $\Psi$ propagates in the ``$-z$''
direction. In the same way, the field ${\Chi}$ is expanded as
\begin{equation}
\label{chiexpansion}
{\Chi}=\int_{0}^{\infty}{\frac{\ud k}{2\pi2k}\left[d^{\dag}(k) \ue^{ik(t-z)}
 + \underline{d}(k) \ue^{-ik(t-z)} \right] Y}.
\end{equation}
The Fourier transform will be written with the normalization
convention
\begin{equation}
\label{delta}
\int{\ud z\, \ue^{i(k-k^\prime)z}} = 2\pi \delta(k-k^\prime).
\end{equation}

\subsubsection{Creation and annihilation operators}

The Fourier coefficients can be expressed as functions of the
spinor field $\Psi$ or $\Psi^\dag$. With equation
(\ref{psiexpansion}) and (\ref{delta}), let us compute the
following integral
\begin{equation}
\int{r \,\ud r\, \ud \theta \,\ud z\, \ue^{ik(t+z)} X^\dag \Psi}=
\int{\frac{\ud k^\prime}{2k^\prime}
\left[b^\dag(-k^\prime)\delta(k+k^\prime) + b(-k^\prime) \delta(k-k^\prime)
\right]} \|X\|^2 =
\frac{\|X\|^2}{2k} b(-k),
\end{equation}
where we have defined
\begin{equation}
\|X\|^2 \equiv \int{r\, \ud r \,\ud\theta \, |X|^2}.
\end{equation}
Note that the separation between $b$ and $b^\dag$ only arises
from the chirality of the spinor field because the integration is
performed only over positive values of the momentum $k$; this is
why the $\delta(k+k^\prime)$ term vanishes. In the following we
will assume that the zero modes are normalized to unity,
$\|X\|^2=1$, and $\|Y\|^2=1$. Playing with similar integrals gives
us the other expansion coefficients
\begin{equation}
\begin{array}{ccccccclll}
\label{psioperators}
\underline{b}(-k) & = & 2k \displaystyle \int{r\,\ud r\,\ud \theta \,\ud z\,
\ue^{ik(t+z)} X^\dag \Psi},
& \quad &
\underline{b}^\dag(-k) & = & 2k \displaystyle \int{r\,\ud r\,\ud
\theta \,\ud z\, \ue^{-ik(t+z)} \Psi^\dag X},
\\ \\
b^\dag(-k) & = & 2k \displaystyle \int{r\,\ud r\,\ud \theta \,\ud z\,
\ue^{-ik(t+z)} X^\dag \Psi},
& \quad &
b(-k) & = & 2k \displaystyle \int{r\,\ud r\,\ud \theta \,\ud z\,
\ue^{ik(t+z)} \Psi^\dag X},
\end{array}
\end{equation}
and the corresponding relations for the spinor field ${\Chi}$
\begin{equation}
\begin{array}{ccccccclll}
\label{chioperators}
d^\dag(k) & = & 2k \displaystyle \int{r\,\ud r\,\ud \theta \,\ud z\,
\ue^{-ik(t-z)} Y^\dag {\Chi}},
& \quad &
d(k) & = & 2k \displaystyle \int{r\,\ud r\,\ud \theta \,\ud z\,
\ue^{ik(t-z)} {\Chi}^\dag Y},
\\ \\
\underline{d}(k) & = & 2k \displaystyle \int{r\,\ud r\,\ud \theta
\,\ud z\, \ue^{ik(t-z)} Y^\dag {\Chi}},
& \quad &
\underline{d}^\dag(k) & = & 2k \displaystyle \int{r\,\ud r\,\ud \theta
\,\ud z\, \ue^{-ik(t-z)} {\Chi}^\dag Y}.
\end{array}
\end{equation}
{}From these, one gets the necessary relations to define creation
and annihilation operators
\begin{equation}
\begin{array}{ccccclll}
\underline{b}^\dag(-k) = \left[\underline{b}(-k)\right]^\dag,
& \quad & \textrm{and} & \quad &
b^\dag(-k) = \left[b(-k)\right]^\dag,
\\ \\
d^\dag(k) = \left[d(k)\right]^\dag,
& \quad & \textrm{and} & \quad &
\underline{d}^\dag(k) = \left[\underline{d}(k)\right]^\dag.
\end{array}
\end{equation}

\subsubsection{Commutation relations}

The canonical quantization is performed by the transformation of
Poisson brackets into anticommutators. Here, we want to quantize
the spinor fields only along the string, and therefore let us
postulate the anticommutation rules \emph{at equal times} for the
quantum fields
\begin{eqnarray}
\label{psiquantization}
\left\{\Psi_\alpha(t,\vec x), \Psi^{\dag \beta}
(t,\vec x^\prime) \right\} & = & \delta(z-z^\prime) X_{\alpha}(r,
\theta) X^{\dag \beta}(r^\prime, \theta^\prime),
\\
\label{chiquantization}
\left\{{\Chi}_\alpha(t,\vec x), {\Chi}^{\dag \beta}
(t,\vec x^\prime) \right\} & = & \delta(z-z^\prime) Y_{\alpha}(r,
\theta) Y^{\dag \beta}(r^\prime, \theta^\prime),
\end{eqnarray}
with $\alpha$ and $\beta$ the spinorial indices, and all the other
anticommutators vanishing. With equation (\ref{psioperators}) and
these anticommutation rules, it follows immediately, for creation
and annihilation operators, that
\begin{equation}
\label{creatoranticom}
\begin{array}{ccccclll}
\left\{b(-k), b^\dag(-k^\prime) \right\} & = & \left\{\underline{b}(-k),
\underline{b}^\dag(-k^\prime)
\right\} & = & 2\pi2k2k^\prime \delta(k-k^\prime),
\\ \\
\left\{\underline{d}(k), \underline{d}^\dag(k^\prime)\right\} & = &
\left\{d(k), d^\dag(k^\prime) \right\} & = & 2\pi2k2k^\prime
\delta(k-k^\prime),
\end{array}
\end{equation}
with all other anticommutators vanishing. From the expressions
(\ref{psiexpansion}) and (\ref{chiexpansion}) and with the
anticommutation rules (\ref{creatoranticom}), it is possible to derive the
anticommutator between two quantum field operators \emph{at any time}. For
instance, the anticommutator between $\Psi$ and $\Psi^\dag$ reads
\begin{eqnarray}
\left\{\Psi_\alpha({\bold{x}}), \Psi^{\dag \beta} ({\bold{x}}^\prime) \right\}
& = &\int{\frac{\ud k\, \ud k^\prime}{(2\pi)^2 2k 2k^\prime}
\left\{b^\dag(-k)\ue^{ik(t+z)} + \underline{b}(-k)\ue^{-ik(t+z)} , b(-k^\prime)
\ue^{-ik^\prime (t^\prime+z^\prime)} \right. } + \nonumber \\
& + & \left. \underline{b}^\dag(-k^\prime) \ue^{ik^\prime (t^\prime+z^\prime)}
\right\} X_\alpha(r,\theta) X^{\dag \beta}(r',\theta').
\end{eqnarray}
Thanks to the delta function coming from the anticommutators
between the $b$ and $b^\dag$, this expression reduces to
\begin{eqnarray}
\left\{\Psi_\alpha({\bold{x}}), \Psi^{\dag \beta} ({\bold{x}}^\prime) \right\}
 & = & i \partial_t \Delta(t-t'+ z-z') X_\alpha(r,\theta)
X^{\dag \beta}(r',\theta'),
\end{eqnarray}
with $\Delta({\bold{x}})$ the well known Pauli-Jordan function which
vanishes for spacelike separation, so the spinor fields indeed respect
micro-causality along the string.

\subsubsection{Fock states}

The Fock space can be built by application of the creation
operators on the vacuum state $|\emptyset\rangle$ which by
definition has to satisfy
\begin{equation}
\label{vide}
\underline{b}(-k)|\emptyset\rangle = b(-k)|\emptyset\rangle =
d(k)|\emptyset \rangle = \underline{d}(k)|\emptyset\rangle = 0,
\end{equation}
and is normalized to unity, i.e., $\langle \emptyset|\emptyset
\rangle = 1$. Each Fock state represents one possible combination
of the fields exitation levels. Let $|{\mathcal{P}} \rangle$ be a
Fock state representing $N_\psi$ particles labeled by $i$ and
$\overline{N}_\psi$ anti-particles labeled by $j$, of kind
$\Psi$, with respective momenta $k_i$ and $l_j$, and, $N_\chi$
particles labeled by $p$ and $\overline{N}_\chi$ anti-particles
labeled by $q$, of kind ${\Chi}$, with respective momenta $r_p$
and $s_q$. By construction the state is
\begin{eqnarray}
\label{fockstate}
|{\mathcal{P}} \rangle & = & b^\dag(-k_1) \ldots b^\dag(-k_i) \ldots
b^\dag(-k_{N_\psi}) \underline{b}^\dag(-l_1) \ldots
\underline{b}^\dag(-l_j) \ldots
\underline{b}^\dag(-l_{\overline{N}_\psi}) d^\dag(r_1) \ldots
\nonumber
\\
& \ldots & d^\dag(r_p) \ldots d^\dag(r_{N_\chi})
\underline{d}^\dag(s_1) \ldots \underline{d}^\dag(s_q) \ldots
\underline{d}^\dag(s_{\overline{N}_\chi}) |\emptyset\rangle.
\end{eqnarray}
Normalizing such a state is done thanks to the anticommutators
(\ref{creatoranticom}). For instance, for a one particle $\Psi$ state
with $k$ momentum, using Eq.~(\ref{vide}), the orthonormalization of
the corresponding states reads
\begin{equation}
\langle k'|k \rangle = 2\pi 2k2k'\delta(k-k').
\end{equation}
Obviously similar relations apply to all the other particle and anti-particle
states. Keeping in mind that the observable values of quantum operators are
their eigenvalues in a given quantum state, let us compute the average of the
occupation number operator involved in many quantum operators, as will be
shown. For $\Psi$ particles it is
\begin{equation}
\label{twoaverage}
\frac{\langle {\mathcal{P}}|b^\dag(-k) b(-k')|{\mathcal{P}}\rangle}{\langle
{\mathcal{P}}|{\mathcal{P}} \rangle} = \frac{2\pi}{\delta(0)}2k2k' \sum_i
\delta(k-k_i)\delta(k'-k_i)
\end{equation}
and analogous relations for the other particle and anti-particle
states. By definition of the Fourier transform (\ref{delta}), the
infinite factor $\delta(0)$ is simply an artifact related to the
length of the string $L$ by
\begin{equation}
L = 2\pi \delta(0),
\end{equation}
in the limit where this string length $L\to \infty$. Note that
using periodic boundary conditions on $L$ allows to consider
large loops with negligible radius of curvature.

\subsection{Fermionic energy momentum tensor}

The simplest way to derive an energy momentum tensor already
symmetrized is basically from the variation of the action with
respect to the metric. Moreover, the Hamiltonian density of the
fermion $\Psi$,
\begin{equation}
{\mathcal{H}}_\psi=\partial_t \Pi_\psi \Psi + \partial_t \overline{\Psi}
\overline{\Pi}_{\psi} - {\mathcal{L}}_\psi,
\end{equation}
with $\Pi$ the conjugate field $\Pi=i \overline{\Psi}\gamma^0$, is also equal
to the $T_\psi^{tt}$ component of the stress tensor. In our case the metric is
cylindrical and we assume a flat Minkowski space-time background, thus in the
fermionic sector the stress tensor reads
\begin{eqnarray}
\label{tmunu}
T_\psi^{\mu \nu} = 2\frac{\delta{\mathcal{L}}_\psi}{\delta g_{\mu \nu}} -
g^{\mu \nu} {\mathcal{L}}_\psi
\quad & \textrm{and} & \quad
T_\chi^{\mu \nu} = 2\frac{\delta{\mathcal{L}}_\chi}{\delta g_{\mu \nu}} -
g^{\mu \nu} {\mathcal{L}}_\chi.
\end{eqnarray}
Once again, let us focus on $\Psi$. Plugging Eq.~(\ref{psilagrangian})
into Eq.~(\ref{tmunu}) gives
\begin{equation}
\label{psitensor}
T_\psi^{\mu \nu} = \frac{i}{2} \overline{\Psi} \gamma^{(\mu}
\partial^{\nu)}
\Psi - \frac{i}{2} \left(\partial^{(\mu}\overline{\Psi}\right)
\gamma^{\nu)} \Psi - B^{(\mu} j^{\nu)}_{_\psi}.
\end{equation}

\subsubsection{Symmetrized Hamiltonian}
\label{toyvac}
{}From Noether theorem, the Hamiltonian $P^t$ is also given by the
conserved charge associated with the time component of the
energy momentum tensor
\begin{equation}
T^{tt}_\psi=i \overline{\Psi}\gamma^0 \partial_t \Psi - i\left(\partial_t
\overline{\Psi} \right) \gamma^0 \Psi.
\end{equation}
Thanks to the expression of the quantum fields in equations
(\ref{psiexpansion}) and (\ref{chiexpansion}), and using the
properties of the zero modes from Eq.~(\ref{currentszeromodes}), the
quantum operator associated to $T_\psi^{tt}$ reads
\begin{eqnarray}
\label{psitensortt}
T_\psi^{tt} & = & \frac{1}{2} \int{\frac{\ud k\,\ud k'}{(2\pi)^22k}}
\left[\left(-b(-k)b^\dag(-k') + \underline{b}^\dag(-k')\underline{b}(-k)
\right) \ue^{i(k'-k)(t+z)} \right.
\nonumber \\
& + &
\left.
\left(\underline{b}^\dag(-k)\underline{b}(-k') -b(-k') b^\dag(-k)
\right) \ue^{-i(k'-k)(t+z)} \right.
\nonumber \\
& + &
\left.
\left(b(-k)\underline{b}(-k')-b(-k')\underline{b}(-k)\right)
\ue^{-i(k'+k)(t+z)} \right.
\nonumber \\
& + &
\left.
\left(-\underline{b}^\dag(-k) b^\dag(-k')+ \underline{b}^\dag(-k')b^\dag(-k)
\right)\ue^{i(k'+k)(t+z)} \right] |X|^2.
\end{eqnarray}
The Hamiltonian is given by spatial integration of the Hamiltonian
density, or similarly from Eq.~(\ref{psitensortt}),
\begin{equation}
\label{hamiltonian}
P^t_\psi= \int{\frac{\ud k}{2\pi2k} \left[-b(-k) b^\dag(-k)
+\underline{b}^\dag(-k) \underline{b}(-k) \right]}.
\end{equation}
Note, once again, that all the terms in the form $b^\dag b^\dag$
or $b{}b$ vanish as a consequence of the chiral nature of the
spinor fields which only allows $k>0$. The average value of this
Hamiltonian in the vacuum is not at all positive, but a simple
renormalization shift is sufficient to produce a reasonable
Hamiltonian provided one uses fermionic creation and annihilation
operators with the corresponding definition for the normal ordered
product (antisymmetric form). The normal ordered Hamiltonian is
therefore well behaved and reads
\begin{eqnarray}
\label{normorder}
:P^t_\psi: & = & \int{\frac{\ud k}{2\pi2k} \left[b^\dag(-k) b(-k)
+ \underline{b}^\dag(-k) \underline{b}(-k) \right]}.
\end{eqnarray}
However, note that such a normal ordering prescription overlooks the
differences between the vacuum energy of empty space, and that in the
presence of the string for the massless fermions. Formally, from
Eq.~(\ref{psitensortt}), the normal ordered Hamiltonian is also obtained
by adding the operator $\int{\ud \vec x \, \uV_\psi}$ to the infinite
Hamiltonian in Eq.~(\ref{hamiltonian}), with
\begin{eqnarray}
\uV_\psi=:T^{tt}_\psi: \, - \, T^{tt}_\psi & = &
\frac{1}{2} \int{\frac{\ud k\,\ud k'}{(2\pi)^22k}}
\left[\left\{ b(-k),b^\dag(-k') \right\} \ue^{i(k'-k)(t+z)} \right.
\nonumber \\
& + &
\left.
\left\{b(-k'), b^\dag(-k) \right\} \ue^{-i(k'-k)(t+z)} \right]|X|^{2}.
\end{eqnarray}
Owing to the anticommutation rules in Eq.~(\ref{psiquantization}), this
expression reduces to
\begin{equation}
\label{inftystruct}
\uV_\psi=\int{\frac{\ud k}{\pi}\, k \,|X|^2}.
\end{equation}
This infinite renormalizing term of the vacuum associated with
the zero modes on the string comes from the contribution of the
infinite renormalization of the usual empty space together with a
finite term representing the difference between the two kinds of
vacua. The previous expression (\ref{inftystruct}) emphasizes the
structure of the divergence, and it can be conjectured that the
finite part is simply obtained by a cut-off $\Lambda$ in momentum
values. The finite vacuum contribution to the stress tensor can
therefore be represented from Eq.~(\ref{inftystruct}), up to the
sign, by the energy density
\begin{eqnarray}
\label{vacuum}
\frac{\Lambda^2}{2\pi}|X|^2=\frac{2\pi}{L_v^2}|X|^2,
& \quad \textrm{with} \quad &
L_v=\frac{2\pi}{\Lambda}.
\end{eqnarray}
The precise determination of the value of $L_v$ is outside the
scope of this simple model. It is well known however, that the
vacuum effects generally involve energies smaller than the first
quantum energy level and consequently it seems reasonable to
consider that $L_v > L$.
For a large loop, $L_v$ can be roughly estimated using the
discretization of the momentum values. With $k_n=2\pi n/L$,
$\uV_\psi$ therefore reads
\begin{equation}
\uV_\psi = \frac{4 \pi}{L^2}|X|^2\sum_{n=0}^{\infty}n.
\end{equation}
Assuming that the vacuum associated with the fermionic zero modes
on the string matches the Minkowski one associated with massless
fermionic modes, in the infinite string limit~\cite{fulling,kay},
$L_v$ can be obtained by substracting the two respective
values of $\uV_\psi$, once the transverse coordinates have been
integrated over.
The infinite sum over $n$ can be regularized by a cut-off factor
$\ue^{-\varepsilon k_n}$, letting $\varepsilon$ equal to zero at the
end of the calculation~\cite{birrell}. The regularized expression of
$\uV_\psi$ finally reads
\begin{equation}
\uV_\psi=\frac{4 \pi}{L^2} |X^2| \sum_{n=0}^{\infty} n
\ue^{-\varepsilon\frac{2\pi}{L}n},
\end{equation}
and expanded asymptotically around $\varepsilon=0$, it yields, once
the transverse coordinates have been integrated,
\begin{equation}
\label{devvac}
\int{r \, \ud r \, \ud \theta \, \uV_\psi} \sim \frac{1}{\pi
\varepsilon^2}- \frac{\pi}{3L^2}.
\end{equation}
As a result, the infinite renormalizing term relevant with the
usual vacuum associated with two dimensional chiral waves
is just $1/\pi \varepsilon^2$ whereas the relevant vacuum
associated with the zero modes along the string is exactly
renormalized by $\int{r \, \ud r \, \ud \theta \, \uV_\psi}$ given in
Eq.~(\ref{devvac}), and therefore involves the finite term $\pi/3L^2$
with a minus sign. As a result, the string zero mode vacuum appears
as an exited state in the Minkowski vacuum associated with two
dimensional chiral modes, with positive energy density
$\pi/3L^2$. In this case, the short distance cut-off therefore
reads
\begin{equation}
L_v^2=6L^2.
\end{equation}
It is therefore necessary to add the $2\pi L/L_v^2$ term to the
canonical normal ordered prescription, in Eq.~(\ref{normorder}), in
order to obtained an Hamiltonian with zero energy for the
Minkowski vacuum associated with two dimensional chiral modes.

It is important to note that, in order to be consistent, the
previous calculations can only involve the vacuum associated with
the corresponding quantized modes, i.e. in this case the zero modes.
Therefore this does not take care of the massive modes. In fact, even
for zero occupancy of the massive bound states, the physical vacuum
along the string must involve a similar dependence in the vacuum
associated with the massive modes. The influence of the two
dimensional massive vacuum on the equation of state will be more
discussed in section \ref{enertens}.

\subsubsection{Stress tensor}

All the other terms of the energy momentum tensor can be derived
from equation (\ref{tmunu}). From the relationships verified by
the zero modes currents in Eq.~(\ref{currentszeromodes}), all the
transverse kinetic terms vanish. Moreover, the only non-vanishing
components of the axial and vectorial currents (\ref{currents})
are $j^t_\psi$ and $j^z_\psi$. Finally, the energy momentum
tensor reads
\begin{equation}
T^{\mu \nu}_\psi = \left(
\begin{array}{ccccllll}
T_\psi^{tt} & 0 & T^{t \theta}_\psi & T^{tz}_\psi \\
0 & 0 & 0 & 0 \\
T^{t \theta}_\psi & 0 & 0 & T^{z \theta}_\psi \\
T^{tz}_\psi & 0 & T^{z \theta}_\psi & T^{zz}_\psi
\end{array}
\right).
\end{equation}
Because the zero modes are eigenvectors of $\gamma^0 \gamma^3$, the
operators $\gamma^0 \partial^0$ and $\gamma^3 \partial^3$ are formally
identical for each spinor field, and therefore the diagonal terms of the
stress tensor are identical
\begin{equation}
\label{eigenstates}
T^{zz}_\psi=T^{tt}_\psi.
\end{equation}
{}From equation (\ref{tmunu}), we find the transverse terms to be
$T^{t \theta}_\psi = - B^\theta j^t_\psi$, and $T^{z \theta}_\psi
= -B^\theta j^z_\psi$. In a Cartesian basis, these components
yield the transverse terms $T^{tx}_\psi$, $T^{ty}_\psi$,
$T^{zx}_\psi$, and $T^{zy}_\psi$, which vanish once the
transverse degrees of freedom have been integrated over. The only
non-vanishing non-diagonal part of the stress tensor comes from
the lightlike nature of each fermion current and reads
\begin{equation}
T^{tz}_\psi=-2i \overline{\Psi} \gamma^0 \partial_t \Psi,
\end{equation}
while the counterpart of the field $\Chi$ gets a minus sign because
of its propagation in the ``$+z$''-direction,
\begin{equation}
T^{tz}_\chi=2i \overline{\Chi} \gamma^0 \partial_t \Chi.
\end{equation}
In the above expressions, the backreaction is neglected, but the
fermionic current, $j^{\mu}=j^{\mu}_\psi+j^{\mu}_\chi$,
generates,  from Eq.~(\ref{gaugemvt}), new gauge field components,
$B_t$ and $B_z$, which have to be small, compared to the
orthoradial component, $B_\theta$, in order to avoid significant
change in the vortex background. However, up to first order, they
provide corrections to the energy momentum tensor whose effects
on energy per unit length and tension will be detailed, in the
classical limit, in section \ref{etat}. The backreaction
correction to the two-dimensional stress tensor then reads
\begin{equation}
\label{brcorr}
T^{\alpha \beta}_{b.r.}= \left(
\begin{array}{ccccllll}
\displaystyle{-2B_t j^t + \frac{1}{2} (\partial_r B_t)^2
+ \frac{1}{2} (\partial_r B_z)^2}
& \displaystyle{ B_z j^t-B_t j^z - (\partial_r B_t)(\partial_r B_z)} \\
\displaystyle{B_z j^t-B_t j^z - (\partial_r B_t)(\partial_r B_z)} &
\displaystyle{2B_z j^z+ \frac{1}{2} (\partial_r B_t)^2
+ \frac{1}{2} (\partial_r B_z)^2}
\end{array}
\right).
\end{equation}

\subsection{Axial and vectorial currents}

From the expression of the spinor fields, the current operators
are immediately found in the Fock space. Moreover, it is
interesting to compute the electromagnetic-like fermionic
current, its scalar analogue being involved in the equation of
state for a cosmic string with bosonic current carriers
\cite{neutral}. From an additional \emph{global} $U(1)$
invariance of the Lagrangian, the electromagnetic-like current
takes similar form as the vectorial one coupled to the string
gauge field. It physically represents the neutral limit of the
full electromagnetic coupling.

\subsubsection{Vectorial currents}

Let $J^\mu$ be the electromagnetic current in the neutral limit.
{}From Noether theorem with global $U(1)$ invariance, this is
\begin{equation}
J^{\mu} = J^\mu_\psi+J^\mu_\chi=- \overline{\Psi} \gamma^{\mu} \Psi +
\overline{{\Chi}} \gamma^\mu {\Chi},
\end{equation}
The fermions $\Psi$ and $\Chi$ carry opposite
electromagnetic-like charges in order to cancel anomalies
\cite{witten}. Using equation (\ref{psiexpansion}), their
components read
\begin{equation}
\begin{array}{cccccll}
\label{empsicurrent}
:J^t_\psi: & = & -:J^z_\psi: & = &  :{\mathcal{J}}_\psi: |X|^2,
\\
\label{emchicurrent}
:J^t_\chi: & = & :J^z_\chi: & = &  - :{\mathcal{J}}_\chi: |Y|^2,
\end{array}
\end{equation}
with the quantum operators defined as
\begin{eqnarray}
:{\mathcal{J}}_\psi: & = & \int{\frac{\ud k\,\ud k'}{(2\pi)^2 2k2k'}}
\left[-b^\dag(-k')b(-k) \ue^{i(k'-k)(t+z)} + \underline{b}^\dag(-k)
\underline{b}(-k') \ue^{-i(k'-k)(t+z)}+ \right.
\nonumber \\
& + & \left. b(-k)\underline{b}(-k') \ue^{-i(k+k')(t+z)} + \underline{b}^\dag(-k)
b^\dag(-k') \ue^{i(k+k')(t+z)} \right] |X|^2,
\\
:{\mathcal{J}}_\chi: & = & \int{\frac{\ud k\,\ud k'}{(2\pi)^2 2k2k'}}
\left[-d^\dag(k')d(k) \ue^{i(k'-k)(t-z)} + \underline{d}^\dag(k)
\underline{d}(k') \ue^{-i(k'-k)(t-z)} + \right.
\nonumber \\
& + & \left. d(k)\underline{d}(k') \ue^{-i(k+k')(t-z)} + \underline{d}^\dag(-k)
d^\dag(k') \ue^{i(k+k')(t-z)} \right] |Y|^2.
\end{eqnarray}
The conserved charges carried by these currents are basically derived from
spatial integration of the corresponding current densities, and are
\begin{eqnarray}
:Q_\psi: & = &  \int{\frac{\ud k}{2\pi (2k)^2} \left[-b^\dag(-k)b(-k)
+ \underline{b}^\dag(-k) \underline{b}(-k) \right]},
\\
:Q_\chi: & = & - \int{\frac{\ud k}{2\pi (2k)^2} \left[-d^\dag(k)d(k)
+ \underline{d}^\dag(k) \underline{d}(k) \right]}.
\end{eqnarray}
As expected at the quantum level, the anti-particles carry
charges that are opposite to that of the particles for both fields.
This is again because the opposite chirality of the fields makes
the $b{}b$ and $b^\dag b^\dag$ terms vanishing. The vectorial
gauge currents are easily obtained from the neutral limit ones
replacing the electromagnetic charge by the $U(1)$ gauge one,
namely
\begin{equation}
\label{vectorialcurrents}
\begin{array}{lllll}
\vspace{4pt} \displaystyle :j^t_{\psi_V}: & = & :-j^z_{\psi_V}: &
= & q \displaystyle{\frac{\cpsir+ \cpsil}{2}} :{\mathcal{J}}_\psi:
|X|^2,
\\
\displaystyle :j^t_{\chi_V}: & = & :j^z_{\chi_V}: & = & q
\displaystyle{\frac{\cchir+ \cchil}{2}} :{\mathcal{J}}_\chi:
|Y|^2.
\end{array}
\end{equation}

\subsubsection{Axial currents}

In the same way, the axial currents are derived from their
classical expressions as function of the quantum fields, from
Eq.~(\ref{currentszeromodes}),
\begin{equation}
\label{axialcurrents}
\begin{array}{lllll}
\vspace{4pt} \displaystyle :j^t_{\psi_A}: & = & - :j^z_{\psi_A}:
& = & q \displaystyle{\frac{\cpsir- \cpsil}{2}}
:{\mathcal{J}}_\psi: \left(|\xi_2|^2-|\xi_3|^2 \right),
\\
\displaystyle :j^t_{\chi_A}: & = & :j^z_{\chi_A}: & = & q
\displaystyle{\frac{\cchir- \cchil}{2}} :{\mathcal{J}}_\chi:
\left(|\xi_1|^2-|\xi_4|^2 \right).
\end{array}
\end{equation}

Thanks to the normalizable zero modes in the transverse plane of the
string, it is possible to construct a Fock space along the string.
The chirality of each spinor field being well defined, anti-particle
states appear at quantum level as another mode propagating at the
speed of light in the same direction than the particle mode, but
carrying opposite gauge and electromagnetic-like charges.
The observable values of the quantum operators previously defined are
given by their average value in the corresponding Fock state. In
particular, the energy per unit length, the tension, and the current
per unit length, can now be derived from the previous expressions.

\section{Equation of state}
\label{etat}

In the case of a scalar condensate in a cosmic string, it was
shown by Peter~\cite{neutral} that the classical formalism of
Carter~\cite{formal} with one single state parameter could apply
and an equation of state for the bosonic cosmic string could be
derived in the form
\begin{equation}
\label{scalareos}
U-T=\sqrt{|w|} {\mathcal{C}},
\end{equation}
where $U$ and $T$ are respectively the energy per unit length and
the tension of the string, ${\mathcal{C}}$ is the current density
along the string and $w$ a state parameter which appear as the
conjugate parameter of ${\mathcal{C}}$ by a Legendre
transformation. An analogous relation can be sought for our string
with fermionic current-carriers from the classical energy per
unit length, tension, and current density values.

Consider the fermionic cosmic string in the quantum state
(\ref{fockstate}). The energy per unit length and tension in this
state are basically given by the eigenvalues associated with
timelike and spacelike eigenvectors, respectively, of the average
value in $|{\mathcal{P}}\rangle$ of the energy momentum tensor,
once the transverse coordinates have been integrated over. The
stress tensor is obviously the total energy momentum tensor
\begin{equation}
T^{\mu \nu} = T^{\mu \nu}_g + T^{\mu \nu}_h +
:T^{\mu \nu}_\psi: + :T^{\mu \nu}_\chi:,
\end{equation}
where $ T^{\mu \nu}_g$ and $T^{\mu \nu}_h$ are the gauge and Higgs
contributions which describe the Goto-Nambu string and which
integrated over the transverse plane provides only two opposite
non-vanishing terms
\begin{equation}
\int{r \,\ud r\,\ud \theta \left(T^{tt}_g + T^{tt}_h \right)} =
-\int{r\,\ud r \,\ud \theta \left(T^{zz}_g + T^{zz}_h \right)}
\equiv M^2,
\end{equation}
thus defining the unit of mass $M$.

\subsection{Average values in the Fock state $|{\mathcal{P}}\rangle$}

\subsubsection{Two-dimensional energy momentum tensor}

Now, let us define the energy momentum tensor operator in two
dimensions, $\overline{T}^{\alpha \beta}$ say, once the
transverse coordinates have been integrated over, and where we
have suppressed the corresponding vanishing terms. Therefore,
with $\alpha$ and $\beta$ equal to $t$ or $z$, and neglecting for
the moment the backreaction, it reads
\begin{equation}
\overline{T}^{\alpha \beta}= \left(
\begin{array}{ccll}
M^2 + \int{r \,\ud r\,\ud \theta \left(:T^{tt}_\psi: + :T^{tt}_\chi:
\right)} & \int{r \,\ud r\,\ud \theta \left(:T^{tz}_\psi: +
:T^{tz}_\chi: \right)} \\
\int{r \,\ud r\,\ud \theta \left(:T^{tz}_\psi: + :T^{tz}_\chi:
\right)} & -M^2 + \int{r \,\ud r\,\ud \theta \left(:T^{tt}_\psi:
+ :T^{tt}_\chi: \right)}
\end{array}
\right).
\end{equation}
The average value in the Fock state $|{\mathcal{P}}\rangle$ of
$\overline{T}^{\alpha \beta}$, is immediately obtained from
equations (\ref{twoaverage}) and (\ref{psitensortt})
\begin{equation}
\label{twostress}
\langle \overline{T}^{\alpha \beta} \rangle_{{\mathcal{P}}} =
\frac{\langle {\mathcal{P}}| \overline{T}^{\alpha \beta}|{\mathcal{P}}
\rangle}{ \langle {\mathcal{P}}|{\mathcal{P}}\rangle} =
\left(
\begin{array}{ccll}
M^2 + E_\chip + E_\psip & E_\chip-E_\psip \\
E_\chip-E_\psip & -M^2+ E_\chip+E_\psip
\end{array}
\right),
\end{equation}
with the notations
\begin{eqnarray}
\label{enerchi}
E_\chip & = &\int{r \,\ud r\,\ud \theta \langle :T^{tt}_\chi:
\rangle_{{\mathcal{P}}}}
= \frac{2}{L} \left(\sum_{i=1}^{N_\psi} k_i + \sum_{j=1}^{
\overline{N}_\psi} l_j \right)_{\mathcal{P}},
\\
\label{enerpsi}
E_\psip & = &\int{r \,\ud r\,\ud \theta\langle :T^{tt}_\psi:
\rangle_{{\mathcal{P}}}}
=\frac{2}{L} \left(\sum_{p=1}^{N_\chi} r_p
+ \sum_{q=1}^{\overline{N}_\chi} s_q \right)_{\mathcal{P}}.
\end{eqnarray}
The summations are just over the momentum values taken in each
particle and anti-particle exitation states, so that $E_\chip$ and
$E_\psip$ depend on the quantum state $|{\mathcal{P}}\rangle$.
The $2 \pi \delta(0)$ factor has been replaced by the physical
length $L$ in order to deal only with finite quantities.
Moreover, from the integral expression of the normal ordering
prescription in Eq.~(\ref{vacuum}), the quantum zero mode vacuum
effects appear simply as a shift of the previous expressions,
and the corrected values of the parameters $E_\psip$ and
$E_\chip$ therefore read
\begin{eqnarray}
\label{fullener}
E_{\psip_v}=E_{\psip}+\frac{2\pi}{L^2_v},
& \quad \textrm{and} \quad &
E_{\chip_v}=E_{\chip}+\frac{2\pi}{L^2_v}.
\end{eqnarray}
Note that, from equations (\ref{enerchi}), (\ref{enerpsi}) and
(\ref{fullener}), if $L_v> L$, the vacuum contribution can be
neglected for non-zero exitation states.

\subsubsection{Current densities}

In the same way, the average value of current operators in the
Fock state $|{\mathcal{P}}\rangle$ are basically derived from the
average of the operators $:{\mathcal{J}}_\psi:$ and
$:{\mathcal{J}}_\chi:$
\begin{equation}
\begin{array}{ccc}
\langle :{\mathcal{J}}_\psi: \rangle_{\mathcal{P}} = -N_\psi +
\overline{N}_\psi,
& \quad &
\langle :{\mathcal{J}}_\chi: \rangle_{\mathcal{P}} = -N_\chi +
\overline{N}_\chi.
\end{array}
\end{equation}
Therefore, the electromagnetic-like current per unit length in the neutral
limit becomes, after transverse integration,
\begin{eqnarray}
\langle \overline{J}^t \rangle_{\mathcal{P}}& = & \frac{1}{L}
\left[\left(-N_\psi + \overline{N}_\psi \right) +
\left(N_\chi - \overline{N}_\chi \right) \right] ,
\\
\langle \overline{J}^z \rangle_{\mathcal{P}}& = & \frac{1}{L}
\left[\left(N_\psi - \overline{N}_\psi \right) +
\left(N_\chi - \overline{N}_\chi \right) \right] .
\end{eqnarray}
Averaging the vectorial and axial gauge currents in equations
(\ref{vectorialcurrents}) and (\ref{axialcurrents}) allows a
derivation of the total fermionic gauge current density
\begin{eqnarray}
\label{gaugecurrentt}
\langle j^t \rangle_{\mathcal{P}} & = & \frac{1}{L}
\left[ f_\psi(r)\left(-N_\psi + \overline{N}_\psi \right)
+ f_\chi(r) \left(-N_\chi + \overline{N}_\chi \right) \right],
\\
\label{gaugecurrentz}
\langle j^z \rangle_{\mathcal{P}} & = & \frac{1}{L}
\left[f_\psi(r) \left(N_\psi - \overline{N}_\psi \right)
+ f_\chi(r) \left(-N_\chi + \overline{N}_\chi \right) \right],
\end{eqnarray}
with the radial functions
\begin{eqnarray}
\label{fchi}
f_\chi(r) & = & q \cchir |\xi_1|^2 + q \cchil |\xi_4|^2,
\\\label{fpsi}
f_\psi(r) & = & q \cpsir |\xi_2|^2 + q \cpsil |\xi_3|^2.
\end{eqnarray}
Moreover, note that these currents can be lightlike, spacelike or
timelike according to the number of each particle species trapped in
the string. For instance, the square magnitude of the
electromagnetic-like line density current reads
\begin{equation}
\overline{{\mathcal{C}}}_{\mathcal{P}}^2 = \langle \overline{J}^t
\rangle_{\mathcal{P}}^2 - \langle \overline{J}^z \rangle_{\mathcal{P}}^2
= \left(\frac{2}{L}\right)^2 \left(N_\psi N_\chi +
\overline{N}_\psi \overline{N}_\chi - \overline{N}_\psi N_\chi
- \overline{N}_\chi N_\psi \right).
\end{equation}
As expected, if there is only one kind of fermion, $\Psi$ or
${\Chi}$, which respectively means $N_\chi=\overline{N}_\chi=0$
or $N_\psi=\overline{N}_\psi=0$, the current is lightlike.
However spacelike currents are also allowed from the existence of
anti-particles as they result from simultaneous exitations between
particles of one kind and anti-particles of the other kind (as for
instance $N_\psi \neq 0$ and $\overline{N}_\chi \neq 0$, or
$N_\chi \neq 0$ and $\overline{N}_\psi \neq 0$). Finally,
timelike currents are obtained from simultaneous exitation between
particles or anti-particles of both kind ($N_\chi \neq 0$ and
$N_\psi \neq 0$, or $\overline{N}_\chi \neq 0$ and
$\overline{N}_\psi \neq 0$).

\subsection{Energy per unit length and tension}
\label{enertens}
In the case of a string having a finite length $L$, periodic
boundary conditions on spinor fields impose the discretization of
the momentum exitation values
\begin{eqnarray}
k_i = \frac{2 \pi}{L} n_{\psi_i},  \qquad  l_j = \frac{2 \pi}{L}
\overline{n}_{\psi_j},  \qquad  r_p = \frac{2 \pi}{L} n_{\chi_p},
\qquad s_q = \frac{2 \pi}{L} \overline{n}_{\chi_q},
\end{eqnarray}
where $n_{\psi_i}$, $\overline{n}_{\psi_j}$, $n_{\chi_p}$ and
$\overline{n}_{\chi_q}$ are positive integers given by the particular
choice of a Fock state. From Eq.~(\ref{enerchi}) and
Eq.~(\ref{enerpsi}), the parameters $E_\chip$ and $E_\psip$ therefore
read
\begin{eqnarray}
\label{enerperio}
E_\chip = \frac{4\pi}{L^2} \left(\sum_{p=1}^{N_\chi} n_{\chi_p} +
\sum_{q=1}^{\overline{N}_\chi} \overline{n}_{\chi_q} \right),
& \quad \textrm{and} \quad &
E_\psip=\frac{4\pi}{L^2} \left(\sum_{i=1}^{N_\psi} n_{\psi_i}
+ \sum_{j=1}^{\overline{N}_\psi} \overline{n}_{\psi_j} \right).
\end{eqnarray}
In the preferred frame where the two-dimensional energy momentum
tensor is diagonal, the energy per unit length and the tension
appear as the eigenvalues associated with the timelike and
spacelike eigenvectors, respectively. By means of equation
(\ref{twostress}), they read
\begin{eqnarray}
\label{ugene}
U_{\mathcal{P}}& = & M^2+2\sqrt{E_\chip E_\psip},
\\
\label{tgene}
T_{\mathcal{P}}& = & M^2-2\sqrt{E_\chip E_\psip}.
\end{eqnarray}
Note, first that the line energy density and the tension always
verify~\cite{prep}
\begin{equation}
\label{tracefix}
U+T=2M^2.
\end{equation}
Moreover the zero mode vacuum effects just modify the parameters
$E_\chip$ and $E_\psip$ as in Eq.~(\ref{fullener}), and therefore do
not modify this relationship. On the other hand, massive modes,
because they are not eigenstates of the $\gamma^0 \gamma^3$
operator, yield vacuum effects which certainly do not modify the
time and space part of the stress tensor in the same way, as was the
case for the zero modes [see Eq.~(\ref{eigenstates})]. As a result,
it is reasonable to assume that the massive mode vacuum effects
modify the equation (\ref{tracefix}) by just shifting the right hand
side by a finite amount, of the order $1/L^2$.

\subsubsection{Classical limit for excited strings}

In order to derive classical values for the energy per unit length
and tension, we do not want to specify in what exitation quantum
states the system is. If the string is in thermal equilibrium
with the external medium, it is necessary to perform quantum
statistics. The number of accessible states in the string is
precisely the total number of combinations between the integer
$n_{\psi_i}$, $\overline{n}_{\psi_j}$, $n_{\chi_p}$, and
$\overline{n}_{\chi_q}$, which satisfies equations (\ref{enerchi})
and (\ref{enerpsi}) for fixed values of the stress tensor,
or, similarly, at given $E_\chi$ and $E_\psi$.
A possible representation of such equilibrium is naturally through
the microcanonical entropy
\begin{equation}
S = k_b \ln {\Omega},
\end{equation}
with $\Omega$ the number of accessible states and $k_b$ the
Boltzmann constant. Let $Q(p,N)$ be the well known partition
function $Q$ which gives the number of partitions of the integer
$p$ into exactly $N$ distinct non-zero integers. With
the following integers
\begin{eqnarray}
K_\chi = \frac{L^2}{4\pi} E_\chi,
& \quad \textrm{and} \quad &
K_\psi = \frac{L^2}{4\pi} E_\psi,
\end{eqnarray}
the number of accessible states $\Omega$ reads
\begin{eqnarray}
\Omega & = & \sum_{n_{\chi}= \frac{N_\chi(N_\chi+1)}{2}}^{K_\chi
-\frac{\overline{N}_\chi( \overline{N}_\chi+1)}{2}} Q(n_\chi, N_\chi)
\sum_{\overline{n}_{\chi}= \frac{\overline{N}_\chi(\overline{N}_\chi
+1)}{2}}^{K_\chi-n_\chi} Q(\overline{n}_\chi, \overline{N}_\chi)
\nonumber \\
& \times &
\sum_{n_{\psi}= \frac{N_\psi(N_\psi+1)}{2}}^{K_\psi-
 \frac{\overline{N}_\psi(\overline{N}_\psi+1)}{2}} Q(n_\psi,N_\psi)
\sum_{\overline{n}_{\psi}= \frac{\overline{N}_\psi(
\overline{N}_\psi+1)}{2}}^{K_\psi-n_\psi} Q(\overline{n}_\psi,
\overline{N}_\psi).
\end{eqnarray}
The energy per unit length and tension of the string will
therefore be the values of $U$ and $T$ which maximize the entropy
at given $N_\psi$, $\overline{N}_\psi$, $N_\chi$ and
$\overline{N}_\chi$. This formalism might be useful whenever
one wants to investigate the dynamics of the string when the
massless current forms, i.e., near the phase transition at high
temperatures. In what follows, we shall assume that whatever the
mechanism through which the fermions got trapped in the string,
they had enough time to reach an equilibrium state with vanishing
temperature. This can be due for instance by a small effective
coupling with the electromagnetic field opening the possibility
of radiative decay~\cite{prep}. If no such effect is present, then
one might argue that the string is frozen in an exited state, the
temperature of which possibly playing the role of a state
parameter~\cite{formal} for a macroscopic description~\cite{bcpc}.

Note that for a given distribution such as those we will be
considering later, the occupation numbers at zero temperature
must be such that, owing to Pauli exclusion principle, the
interaction terms implying for instance a $\Psi\bar\Psi$ decay
into a pair $\Chi\bar\Chi$ through Higgs of $B_\mu$ exchange are
forbidden (vanishing cross-section due to lack of phase space).
In practice, this means that the following analysis is meaningful
at least up to one loop order.

\subsubsection{String at zero temperature}

\begin{figure}
\begin{center}
\epsfig{file=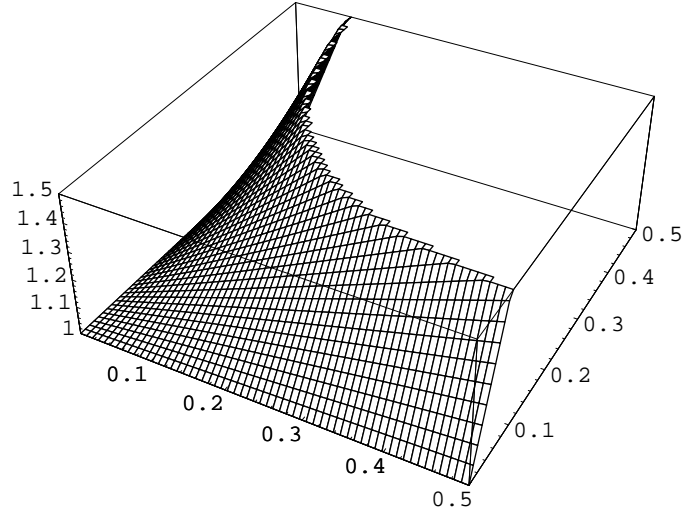,width=9cm}
\end{center}
\caption{The energy per unit length, in unit of $M^2$ and at zeroth order, as
function of $\Chi$ and $\Psi$ fermion densities plotted in unit of $M$, in the
infinite string limit.}
\label{figuzerorder}
\end{figure}

\begin{figure}
\begin{center}
\epsfig{file=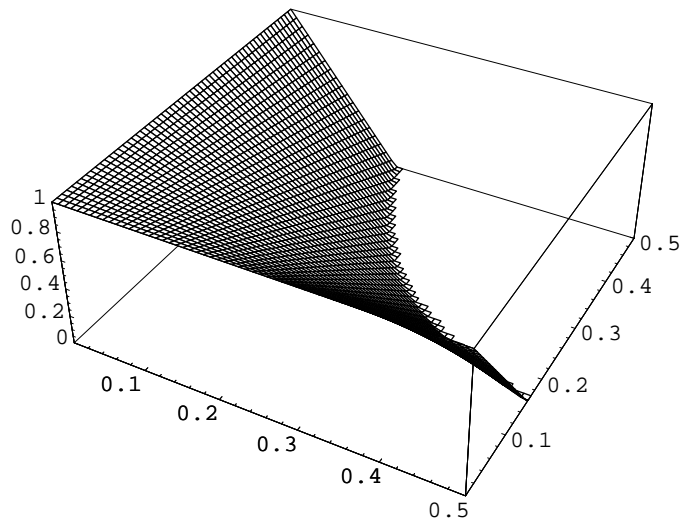,width=9cm}
\end{center}
\caption{The tension, in unit of $M^2$ and at zeroth order, as function of
$\Chi$ and $\Psi$ fermion densities plotted in unit of $M$, in the infinite
string limit.}
\label{figtzerorder}
\end{figure}

\begin{figure}
\begin{center}
\epsfig{file=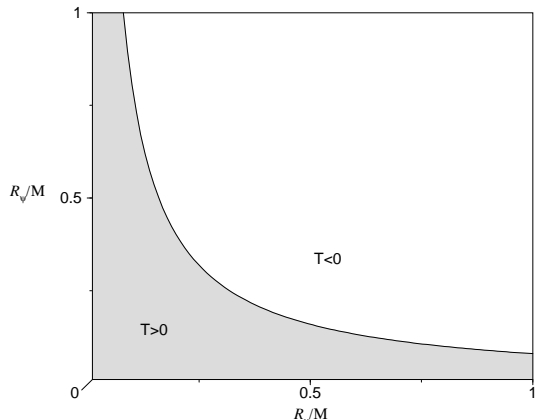,width=7cm}
\caption{Sign of the tension in the $(\Rho_\chi/M,\Rho_\psi/M)$ plane,
in the infinite string limit. According to the macroscopic formalism,
the string is unstable with respect to transverse perturbations for
$T<0$.}
\label{figtzerorderneg}
\end{center}
\end{figure}

For weak coupling between fermions trapped in the string and
external fields, as is to be expected far below the energy scale
where the string was formed, the set of particles is assumed to
fall in the ground state and because of anticommutation rules
(\ref{creatoranticom}) it obeys Fermi-Dirac statistic at zero
temperature. Consequently, the parameters $E_\chip$ and $E_\psip$
reads
\begin{eqnarray}
\label{energround}
E_\psi = \frac{2 \pi}{L} \left[\rho_\psi(L \rho_\psi+1) +
\overline{\rho}_\psi (L \overline{\rho}_\psi+ 1) \right],
& \quad \textrm{and} \quad &
E_\chi = \frac{2 \pi}{L} \left[ \rho_\chi(L \rho_\chi + 1)
+\overline{\rho}_\chi(L \overline{\rho}_\chi + 1) \right],
\end{eqnarray}
where the new parameters $\rho=N/L$ are the line number densities of
the corresponding particles and anti-particles trapped in the string.
Strictly speaking, these are the four independent state parameters
which fully determine the energy per unit length and the tension in
Eq.~(\ref{ugene}) and Eq.~(\ref{tgene}), and so cosmic strings with
fermionic current-carriers do not verify the same equation of state
as the bosonic current-carrier case. This is all the more so manifest
with another more intuitive set of state parameters, $\Rho$ and
$\Theta$, defined for each fermion by
\begin{eqnarray}
\Rho^2  =  \left(\rho+\frac{1}{2L} \right)^2 + \left(\overline{\rho}
+\frac{1}{2L} \right)^2,
& \quad \textrm{and} \quad &
\Theta = \arctan{\left(\frac{\overline{\rho}+\frac{1}{2L}}{\rho
+\frac{1}{2L}}\right)}.
\end{eqnarray}
These are simply polar coordinates in the two-dimensional space defined by the
particle and anti-particle densities of each spinor field. The parameter
$\Rho$ physically represents the fermion density trapped in the string
regardless of the particle or anti-particle nature of the current carriers. It
is therefore the parameter that we would expect to be relevant in a purely
classical approach. On the other hand, $\Theta$ quantifies the asymmetry
between particles and anti-particles since
\begin{eqnarray}
\rho=\Rho \cos{\Theta} - \frac{1}{2L},
& \quad \textrm{and} \quad &
\overline{\rho}=\Rho \sin{\Theta} -\frac{1}{2L}.
\end{eqnarray}
The energy per unit length, tension and line density current now
read
\begin{eqnarray}
\label{uzerorder}
U & = & M^2 + 4\pi\sqrt{\left(\Rho_\chi^2-\frac{1}{2 L^2}\right)
\left(\Rho_\psi^2-\frac{1}{2L^2}\right)},
\\
\label{tzerorder}
T & = & M^2 - 4\pi\sqrt{\left(\Rho_\chi^2-\frac{1}{2 L^2}\right)
\left(\Rho_\psi^2-\frac{1}{2L^2}\right)},
\\
\overline{{\mathcal{C}}}^2 & = & 8 \Rho_\chi \Rho_\psi
\sin{\left(\Theta_\chi-\frac{\pi}{4}\right)}
\sin{\left(\Theta_\psi - \frac{\pi}{4} \right)}.
\end{eqnarray}
There are always four independent state parameters but only two,
$\Rho_\chi$ and $\Rho_\psi$, are relevant for line density energy
and tension. Compared to the scalar case where only one kind of charge
carrier propagates along the string, it is not surprising that we
found two degrees of freedom with two kinds of charge carriers. On the
other hand, the nature of the line density current is not relevant
because it only appears through $\Theta_\psi$ and $\Theta_\chi$, which
not modify $U$ and $T$, at least at the zeroth order. The energy per
unit length and tension relative to $M^2$ are represented in
Fig.~\ref{figuzerorder} and Fig.~\ref{figtzerorder} as function of
$\Rho_\chi/M$ and $\Rho_\psi/M$, in the infinite string limit.
As expected from their analytical expressions in the infinite
string limit, the energy per unit length is always positive and
grows with both fermion densities, $\Rho_\chi$ and $\Rho_\psi$,
whereas the tension always decreases and takes negatives values for
large fermion densities. Obviously, in the case $\Rho_\chi=\Rho_\psi=0$
there is no current along the string and we recover the Goto-Nambu
case, $U=T=M^2$. The chiral case, where the fermionic current is lightlike,
is obtained for $\Rho_\chi=0$, or $\Rho_\psi=0$, and also verifies
$U=T=M^2$ as in the chiral scalar current case~\cite{chiralmodel}.
From Eq.~(\ref{tzerorder}), and in the infinite string limit, the densities
for which the tension vanishes verify
\begin{equation}
\Rho_\chi\Rho_\psi = \frac{M^2}{4 \pi},
\end{equation}
This curve separates the plane $(\Rho_\chi,\Rho_\psi)$ in two
regions where $T$ is positive near the origin, and negative on
the other side (see Fig.~\ref{figtzerorderneg}). In the
macroscopic formalism of Carter~\cite{formal}, the transverse
perturbations propagation speed is given by $c_T^2 = T/U$, and
therefore the domains where $T<0$ correspond to strings which are
always locally unstable with respect to transverse perturbations.
The tension of the string becomes negative only for carrier
densities close to the mass of the Goto-Nambu string
$M$. For such currents, it is necessary to derive the
backreaction in order to see how relevant it is for the energy
per unit length and tension. Moreover, in a renormalizable model,
the vacuum mass acquired by the fermions from their coupling to
the Higgs field is less than the Goto-Nambu string mass, and
thus, another quantum effects may take place before the negative
tension is reached, like tunneling into massive states. Besides,
note that $M$, the string unit of mass, arising from
non-perturbative effects, may well be much larger than the Higgs
mass, and so, it is expected that $\Rho_\chi, \Rho_\psi \ll M$.
Thus, the no-spring conjecture~\cite{nospring} proposed in the
case of bosonic carrier presumably apply to the fermionic carrier
case as well. Moreover, the zero mode vacuum effects on energy per
unit length and tension appear clearly from Eq.~(\ref{fullener}) as
additional string length. The corrected values of the equation of
state are therefore obtained by replacing the physical length of the
string, $L$, in Eq.~(\ref{uzerorder}) and Eq.~(\ref{tzerorder}), by
an equivalent length, $L_{e}$ say, which verifies
\begin{equation}
\label{eqlength}
\frac{1}{2L_e^2}=\frac{1}{2L^2}-\frac{1}{L_v^2}.
\end{equation}
In the particular case where $L_v^2=6L^2$, it reads $L_e^2=(3/2)
L^2$. On the other hand, the massive vacuum effects certainly shift
in a different way $U$ and $T$ by a finite amount as previously
discussed, but will not be considered in the following.
In the next section, the backreaction is derived in
the classical limit in order to find corrected values of energy
per unit length and tension. Moreover we shall take care of the
finite length of the string $L$, keeping in mind that its
value, and consequently the value of $L_e$, have to be larger than
$1/M$ since all physical values have been derived in the classical
vortex background, i.e., the quantum effects of the Higgs field have
been neglected.

\subsection{Backreaction}

The existence of fermionic currents carrying gauge charge along
the string gives rise to new gauge field components, $B_t$ and
$B_z$, from the equations of motion (\ref{gaugemvt}). These, being
coupled with the corresponding currents, provide additional terms
in the energy momentum tensor (\ref{brcorr}). As a first step, the
new gauge field components are computed numerically from the zero
modes solutions of Eq.~(\ref{zeromodes}). The corrected equation of
state is then analytically derived, the numerical dependencies
having been isolated in model dependent coefficients.

\subsubsection{Backreacted gauge fields}

\begin{figure}
\begin{center}
\epsfig{file=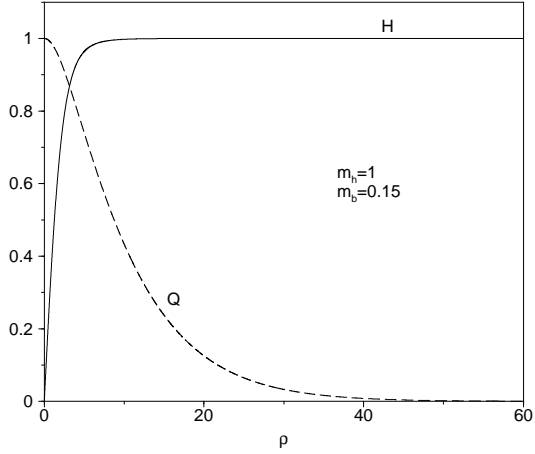,width=7cm}
\caption{The solutions of the field equations for the vortex background. The
Higgs field, $H$, takes its vacuum expectation value at infinity and the
gauge bosons condensate in the vortex.}
\label{figback}
\end{center}
\end{figure}

\begin{figure}
\begin{center}
\epsfig{file=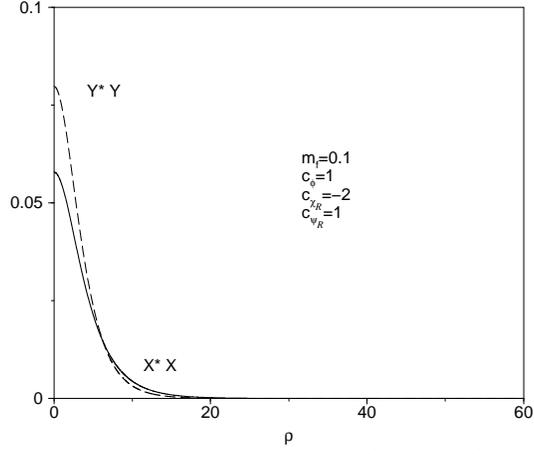,width=7cm} \caption{The normalized
probability densities of the zero modes, $|X|^2$ and $|Y|^2$. The
rapid decay far from the string core reflects the bound state
nature of the condensates.} \label{figproba}
\end{center}
\end{figure}

\begin{figure}
\begin{center}
\epsfig{file=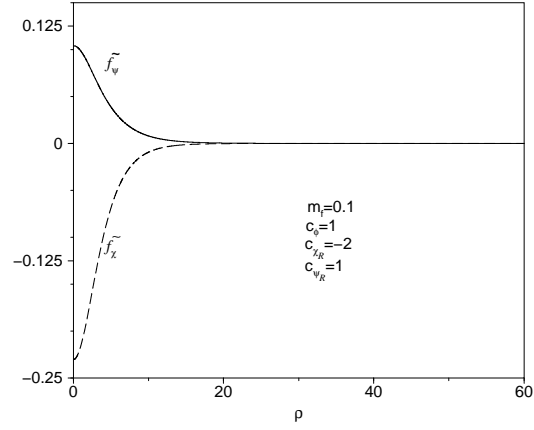,width=7cm}
\caption{The dimensionless radial current functions $\tilde{f}_\chi$ and
$\tilde{f}_\psi$. They can be viewed as the effective transverse density
charge carried by the fermion currents. Their sign results in the initial
choice of each conserved fermion gauge charge.}
\label{figradfunc}
\end{center}
\end{figure}

In order to compute the $B_t$ and $B_z$ fields at first order, we
only need the zeroth order values of the zero modes and the
vortex background. Let us introduce the dimensionless scaled
fields and variables
\begin{eqnarray}
\varphi=\eta H,
\quad
Q_\theta=Q
\quad \textrm{and} \quad
r=\frac{\varrho}{m_h},
\end{eqnarray}
with $m_h=\eta\sqrt{\lambda}$ the classical mass of the Higgs
field. From the equation of motion (\ref{higgsmvt}), the
orthoradial gauge field $Q$ and $H$ are solution of
\begin{eqnarray}
\label{tildehiggs}
\frac{d^2H}{d\varrho^2}+\frac{1}{\varrho} \frac{dH}{d\varrho} & = &
\frac{H Q^2}{\varrho^2}+\frac{1}{2}H(H^2-1),
\\
\label{tildegauge}
\frac{d^2 Q}{d \varrho^2} -\frac{1}{\varrho}\
\frac{d Q}{d \varrho} & = & \frac{m_b^2}{m_h^2}H^2 Q,
\end{eqnarray}
where $m_b=qc_\phi \eta$ is the classical mass of the gauge boson.
The numerical solutions of these equations have been computed
earlier by many people~\cite{bps,neutral} using relaxation methods
\cite{adler}. They are presented in Fig.~\ref{figback} for a
specific (assumed generic) set of parameters.
In the same way, deriving Eq.~(\ref{zeromodes}) with respect to
$\varrho$ yields the right component of the zero mode $Y$ as a
solution of
\begin{eqnarray}
\zeta_1'' - \left(\frac{H'}{H}+\frac{Q-n}{\varrho}\right)\zeta_1'
& + & \left(\frac{\cchir}{\cphi}\left[\frac{Q-n}{\varrho} \frac{H'}{H}
+ \frac{Q-n}{\varrho^2}-\frac{Q'}{\varrho}\right] \right.
\nonumber \\
& - &
\left.
\frac{\cchil \cchir}{\cphi^2} \left(\frac{Q-n}{\varrho}\right)^2
-\frac{m_f^2}{m_h^2} H^2 \right) \zeta_1 = 0,
\end{eqnarray}
while the left one satisfies
\begin{eqnarray}
i \frac{m_f}{m_h}H \zeta_4 & = &  \zeta_1' - \frac{\cchir}{\cphi}
\frac{Q-n}{\varrho} \zeta_1,
\end{eqnarray}
where a prime indicates a derivation with respect to the
dimensionless radial variable $\varrho$. The field $\Psi$
verifies similar equations with the transformation, $\zeta
\rightarrow \xi$ and $c_\chi \rightarrow c_\psi$. The numerical
integration has been performed with a relaxation method~\cite{adler}
and verified on the original system (\ref{zeromodes})
with a shooting method. As a result, the normalized probability
densities of the zero modes, $|X|^2$ and $|Y|^2$, are plotted in
Fig.~\ref{figproba}. The dimensionless radial functions
$\tilde{f}_\chi$ and $\tilde{f}_\psi$ defined from Eq.~(\ref{fchi})
and Eq.~(\ref{fpsi}) by
\begin{equation}
\tilde{f}_\chi=\frac{2 \pi}{m_h^2} f_\chi,
\quad \textrm{and} \quad
\tilde{f}_\psi=\frac{2 \pi}{m_h^2} f_\psi,
\end{equation}
are plotted in Fig.~\ref{figradfunc}. As expected, the fields
are confined in the string core, and so will the corresponding
fermionic currents.

\begin{figure}
\begin{center}
\epsfig{file=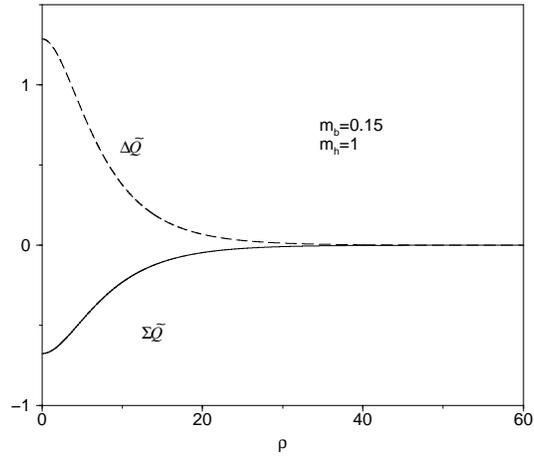,width=7cm}
\caption{The dimensionless backreacted gauge fields $\Delta\tilde{Q}$ and
$\Sigma\tilde{Q}$, generated by the fermion currents. They do not lead
to new long-range effects since they acquire non-zero mass outside the
string core due to their coupling with the Higgs fields.}
\label{figgauge}
\end{center}
\end{figure}

Let us define the more relevant components of the backreacted
gauge field, $\Delta B=B_z-B_t$ and $\Sigma B=B_z+B_t$, with the
corresponding dimensionless scaled fields $\Delta \tilde{Q}$ and
$\Sigma \tilde{Q}$ defined by
\begin{eqnarray}
\label{newgaugetilde}
\Delta B  =  \left(\overline{\rho}_\chi - \rho_\chi \right)
\frac{m_b^2}{\pi \eta^2} \frac{\Delta \tilde{Q}}{q\cphi},
& \quad \textrm{and} \quad &
\Sigma B  =  -\left(\overline{\rho}_\psi - \rho_\psi \right)
\frac{m_b^2}{\pi \eta^2} \frac{\Sigma \tilde{Q}}{q\cphi}.
\end{eqnarray}
The equations of motion (\ref{gaugemvt}) in the classical limit now reads
\begin{eqnarray}
\label{newgaugemvt}
\Delta \tilde{Q}''+\frac{1}{\rho} \Delta\tilde{Q}'-\frac{m_b^2}{m_h^2}H^2
\Delta\tilde{Q}  =  \frac{\tilde{f}_\chi}{q \cphi},
& \quad \textrm{and} \quad &
\Sigma \tilde{Q}''+\frac{1}{\rho} \Sigma \tilde{Q}'-\frac{m_b^2}{m_h^2}H^2
\Sigma \tilde{Q}  =  \frac{\tilde{f}_\psi}{q \cphi}.
\end{eqnarray}
As for fermions, these new gauge fields get their masses from coupling
with the Higgs field, and therefore have non-zero mass outside the string
core. Moreover, they are generated by fermionic massless currents
confined in the core, therefore they also condense in and do not lead
to new long-range effects.
The solutions of these equations (\ref{newgaugemvt}) have been
obtained using, once again, a relaxation method~\cite{adler} and are
represented in Fig.~\ref{figgauge}. Note that owing to the scaled
field $\Delta \tilde{Q}$ and $\Sigma \tilde{Q}$, we have separated
the numerical dependence in the gauge field and currents from the
fermion densities content [see Eq.~(\ref{newgaugetilde})].
Up to now, we have computed the fermionic gauge currents along the
string as well as the component $B_t$ and $B_z$, so that the
backreaction correction to the energy momentum tensor is
computable from Eq.~(\ref{brcorr}). As in the previous section,
the energy per unit length and tension can be derived in the
preferred frame where the stress tensor is diagonal, but now we
have to find the eigenvalues of the full two-dimensional energy
momentum tensor $\overline{T}^{\alpha \beta}_{b.r.}+\langle
\overline{T}^{\alpha \beta} \rangle$, with
\begin{equation}
\overline{T}^{\alpha \beta}_{b.r.}=\int{ r \, \ud r \, \ud \theta
\, T^{\alpha \beta}_{b.r.}}
\end{equation}

\subsubsection{Energy per unit length and tension with backreaction}

Using the dimensionless field, $\Delta \tilde{Q}$ and $\Sigma
\tilde{Q}$, with the expressions of the currents given in equations
(\ref{gaugecurrentt}) and (\ref{gaugecurrentz}), one gets, after
some algebra, the full expression of the stress tensor
with corresponding eigenvalues
\begin{eqnarray}
\label{ufull}
\widehat{U} & = & M^2 -\I(\Theta_\chi,\Theta_\psi)\Rhop_\chi \Rhop_\psi
+4 \pi \sqrt{\left(\Rhop_\chi^2 -\frac{1}{2L_e^2}\right)\left(\Rhop_\psi^2
-\frac{1}{2L_e^2}\right)},
\\
\label{tfull}
\widehat{T} & = & M^2 -\I(\Theta_\chi,\Theta_\psi)\Rhop_\chi \Rhop_\psi
-4 \pi \sqrt{\left(\Rhop_\chi^2 -\frac{1}{2L_e^2}\right)\left(\Rhop_\psi^2
-\frac{1}{2L_e^2}\right)},
\end{eqnarray}
with the scaled state parameters
\begin{eqnarray}
\label{dampdens}
\Rhop_\chi & = & \Rho_\chi \sqrt{1-\frac{m_b^2}{2 \pi^2 \eta^2} \left(
I_{\Delta \chi} - I_{\Delta 2} \right)
\sin^2{\left(\Theta_\chi-\frac{\pi}{4}\right)}},
\\
\Rhop_\psi & = & \Rho_\psi \sqrt{1-\frac{m_b^2}{2 \pi^2 \eta^2} \left(
I_{\Sigma \psi} - I_{\Sigma 2} \right)
\sin^2{\left(\Theta_\psi-\frac{\pi}{4}\right)}},
\end{eqnarray}
and the function $\I(\Theta_\chi,\Theta_\psi)$ defined by
\begin{equation}
\label{ifonction}
\I(\Theta_\chi,\Theta_\psi) = \frac{m_b^2}{\pi \eta^2}
\frac{\left(I_{\Sigma \chi}+I_{\Delta \psi} \right)
\sin{\left(\Theta_\chi-\frac{\pi}{4} \right)}
\sin{\left(\Theta_\psi-\frac{\pi}{4} \right)}}{\sqrt{1-\frac{m_b^2}{2 \pi^2
\eta^2} \left(I_{\Delta \chi}-I_{\Delta 2}\right)
\sin^2{\left(\Theta_\chi-\frac{\pi}{4}\right)}}
\sqrt{1-\frac{m_b^2}{2 \pi^2 \eta^2} \left(I_{\Sigma \psi}-
I_{\Sigma 2}\right) \sin^2{\left(\Theta_\psi-\frac{\pi}{4}\right)}}}.
\end{equation}
The numerical integrations previously carried out appear through
pure numbers which depend only on the model parameters. The
coupling $B_\mu j^\mu$ leads to the following quantities
\begin{eqnarray}
\label{numintegspin}
I_{\Sigma \chi}=-2\int{\,\varrho \, \ud \varrho \, \Sigma \tilde{Q}(\varrho)
\frac{\tilde{f}_\chi(\varrho)}{q\cphi}},
& \qquad &
I_{\Delta \psi}=-2\int{\,\varrho \, \ud \varrho \, \Delta \tilde{Q}(\varrho)
\frac{\tilde{f}_\psi(\varrho)}{q\cphi}},
\\
I_{\Sigma \psi}=-2\int{\,\varrho \, \ud \varrho \, \Sigma \tilde{Q}(\varrho)
\frac{\tilde{f}_\psi(\varrho)}{q\cphi}},
& \qquad &
I_{\Delta \chi}=-2\int{\,\varrho \, \ud \varrho \, \Delta \tilde{Q}(\varrho)
\frac{\tilde{f}_\chi(\varrho)}{q\cphi}},
\end{eqnarray}
while the kinetic contribution of the new gauge fields appears through
\begin{eqnarray}
\label{numintegkin}
I_{\Delta 2}=\int{\, \varrho \, \ud \varrho \, \left(\partial_\varrho
\Delta \tilde{Q}(\varrho) \right)^2},
& \qquad &
I_{\Sigma 2}=\int{\, \varrho \, \ud \varrho \, \left(\partial_\varrho
\Sigma \tilde{Q}(\varrho) \right)^2}.
\end{eqnarray}
By means of the equations of motion (\ref{newgaugemvt}) and the constant
sign of $\Delta \tilde{Q}$ and $\Sigma \tilde{Q}$,
$I_{\Delta \chi}$ and $I_{\Sigma \psi}$ are found to be always positive.
Intuitively, as in electromagnetism, the gauge field generated from charge
currents tends to resist to the currents which give birth to it. In
our case, the backreaction actually damps the weight of the charge
carriers in the energy per unit length and tension. In fact, the
relevant state parameters are now
$\Rhop$ instead of $\Rho$ with $\Rhop<\Rho$ since $I_{\Delta \chi}$ and
$I_{\Sigma \psi}$ are positive. Moreover, numerical calculations show
that the kinetic contribution numbers (\ref{numintegkin}) are always
one order of magnitude smaller than those resulting in the coupling
between gauge fields and currents (\ref{numintegspin}), as expected
for reasonable backreacted gauge field since they only involve the
square gradient of these fields [see Eq.~(\ref{numintegkin})].
However, there is an additional term involving new dependence in
the asymmetry between particles and anti-particles through the
$\I(\Theta_\chi,\Theta_\psi)$ function. In order to understand this
point physically, let us derive the magnitude of the gauge current
carried by the fermions. From Eq.~(\ref{gaugecurrentt}) and
Eq~.(\ref{gaugecurrentz}), once the transverse coordinates have been
integrated over, the dimensionless magnitude reads
\begin{equation}
\jbt^2=2 (2q\cphi)^2
\frac{\tilde{F}_\chi \tilde{F_\psi}\sin{\left(\Theta_\chi-\frac{\pi}{4}
\right)}
\sin{\left(\Theta_\psi-\frac{\pi}{4} \right)}\Rhop_\chi \Rhop_\psi}
{\sqrt{1-\frac{m_b^2}{2 \pi^2
\eta^2} \left(I_{\Delta \chi}-I_{\Delta 2}\right)
\sin^2{\left(\Theta_\chi - \frac{\pi}{4}\right)}}
\sqrt{1-\frac{m_b^2}{2 \pi^2 \eta^2} \left(I_{\Sigma \psi}-I_{\Sigma 2}\right)
\sin^2{\left(\Theta_\psi-\frac{\pi}{4}\right)}}},
\end{equation}
with the dimensionless constants are
\begin{eqnarray}
\tilde{F}_\chi = \int{\, \varrho \, \ud \varrho \,\frac{\tilde{f}_\chi
(\varrho)}{q \cphi}},
& \quad \textrm{and} \quad &
\tilde{F}_\psi = \int{\, \varrho \, \ud \varrho \,\frac{\tilde{f}_\psi
(\varrho)}{q \cphi}}.
\end{eqnarray}
These numbers can be viewed as the effective charge carried by the
fermionic gauge currents since
\begin{equation}
\frac{\jbt^2}{(2q\cphi)^2\tilde{F_\chi}\tilde{F_\psi}}=\frac{\overline{
{\mathcal{C}}}^2}{4}.
\end{equation}
The function $\I(\Theta_\chi,\Theta_\psi)$ therefore verifies
\begin{equation}
\I(\Theta_\chi,\Theta_\psi)\Rhop_\chi \Rhop_\psi
=\frac{I_{\Sigma \chi}+I_{\Delta \psi}}{8 \pi \tilde{F}_\chi \tilde{F}_\psi}
\jbt^2,
\end{equation}
and, as before, according to Eq.~(\ref{newgaugemvt}),
$(I_{\Sigma \chi}+I_{\Delta \psi})/(\tilde{F}_\chi
\tilde{F}_\psi)$ is always positive, so the sign of $\I$ directly
reflects the spacelike or timelike nature of the current. Thus,
in addition to the backreaction damping effect, there is a
correction to the energy per unit length and tension directly
proportional to the magnitude of the fermionic current. Note that
this effect appears as a correction due to backreaction and not,
as it is the case for cosmic string with bosonic current
carriers, at the zeroth order~\cite{neutral}.

\subsubsection{Equation of state with backreaction}

Unfortunately, the corrected expressions of the line density
energy and tension involve four independent state parameters, and
consequently are not easily representable. However, they can be
studied as functions of the damped fermion densities $\Rhop$,
modified only by the function $\I(\Theta_\chi,\Theta_\psi)$ which
quantifies the efficiency of the fermionic currents in generating
backreacted gauge fields. In this way, the comparison with the
zeroth order case is all the more so easy.

The study of the surfaces defined by $\widehat{U}$ and
$\widehat{T}$ in the plane $(\Rhop_\chi,\Rhop_\psi)$ is less
canonical than at zeroth order. Three critical values of
the function $\I$ are found to modify the behaviors of the tension and
energy per unit length, namely, $-4\pi$, $0$, and $4 \pi$. However,
only small values of $\I$ are reasonable in this model as it is
discussed in the next section. This analysis is consequently
constraints to values of $|\I|<4 \pi$.

\paragraph{Energy per unit length.}

The line density energy follows different behaviors according
to the value of $\I$.

\begin{figure}
\begin{center}
\epsfig{file=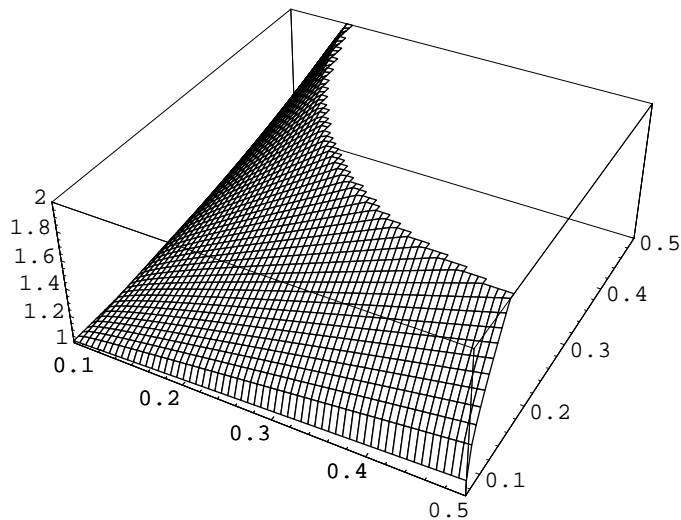,width=9cm}
\caption{The energy per unit length in unit of $M^2$ as function of
$\Rhop_\chi/M$ and $\Rhop_\psi/M$, for spacelike currents with
$\I(\Theta_\chi,\Theta_\psi)<0$. The influence of the finite length
of the string just appears near the axes.}
\label{figuineg}
\end{center}
\end{figure}

The first and simplest case $\I<0$, obtained for spacelike
fermionic gauge currents, is very similar to the zeroth order
case, and the energy per unit length just grows a bit faster
with the damped fermion densities $\Rhop_\chi$ and $\Rhop_\psi$,
as on Fig.~\ref{figuineg}.

\begin{figure}
\begin{center}
\epsfig{file=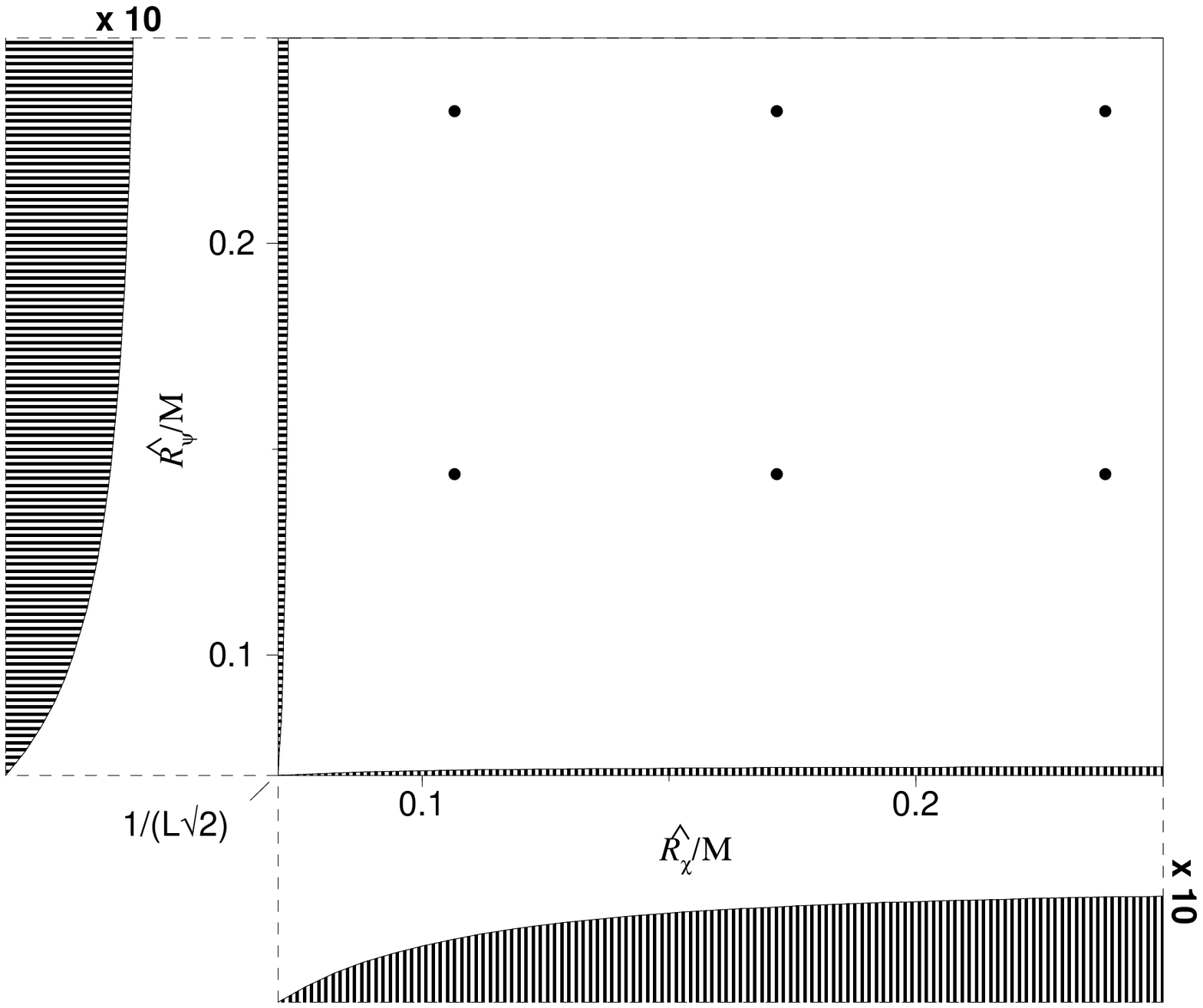,width=7cm}
\epsfig{file=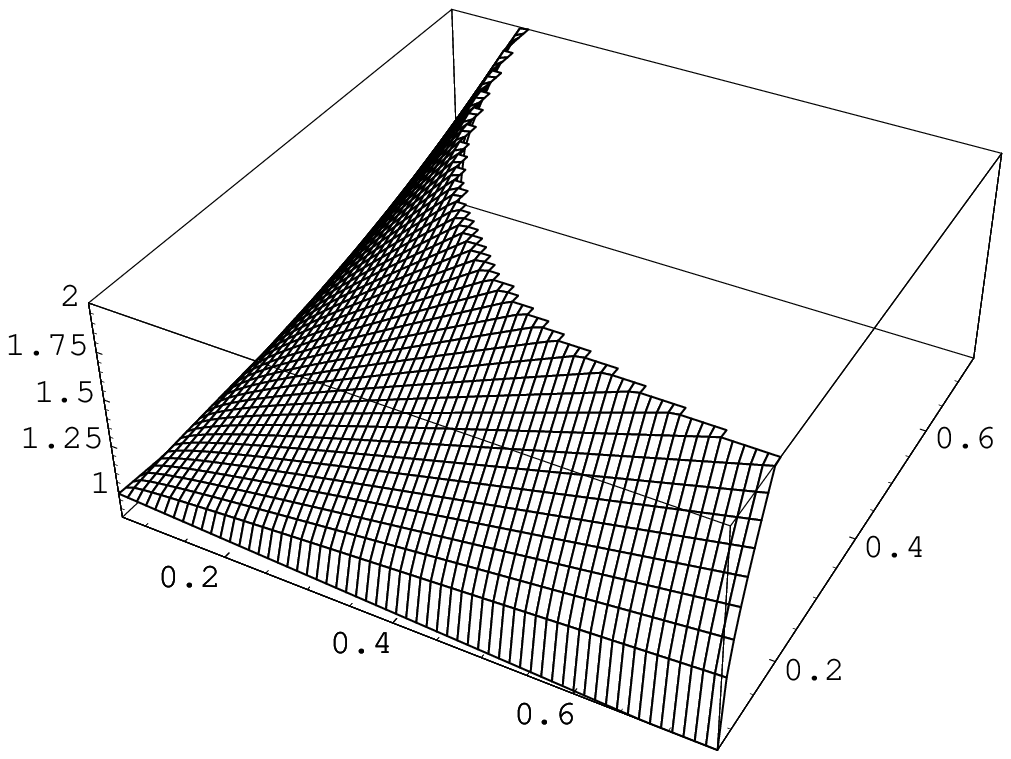,width=9cm}
\caption{Variation domains of the energy per unit length, plotted in
unit of $M^2$, in the $(\Rhop_\chi/M,\Rhop_\psi/M)$ plane for timelike
currents with $0<\I<4\pi$. The vertical hatched regions define domains
where the line density energy could decrease with $\Rhop_\chi$ whereas
the horizontal ones are regions where the line density energy could
decrease with $\Rhop_\psi$. On one hand, these domains are
asymptotically limited by the curves $\Rhop=(2\pi/L_e)
\sqrt{2/(16\pi^2-\I^2)}$ and therefore coincide with the axis for large
values of $L_e$.
On the other hand, for reasonable values of $\I \ll 4\pi$, there is
no accessible quantum state inside, the first one being shown as a
dot. As a result the energy per unit length always grows with the parameters
$\Rhop$ and is always positive.}
\label{figuiposip}
\end{center}
\end{figure}

For timelike currents, $I>0$, we find that the backreaction damps
the growth of the density line energy with the fermion densities. As
a result the line density energy seems to decrease in some regions, and
the stationary curves of $\widehat{U}$ with respect to $\Rhop_\chi$
are given, from Eq.~(\ref{ufull}), by
\begin{eqnarray}
\label{energystat}
\frac{\partial{\widehat{U}}}{\partial \Rhop_\chi}=0 & \Leftrightarrow &
\Rhop_\psi=\frac{2 \pi}{L_e} \Rhop_\chi \sqrt{\frac{2}{\frac{\I^2}{2L_e^2}
+ (16 \pi^2 -\I^2) \Rhop_\chi^2}},
\end{eqnarray}
and thanks to the symmetry between $\Rhop_\chi$ and $\Rhop_\psi$,
similar equations are obtained for $\partial \widehat{U}/\partial
\Rhop_\psi=0$. Finally, the variation domains of the line density
energy are represented in Fig.~\ref{figuiposip} for $0<\I<4\pi$.
The first discrete values of the fermion densities (the length
of the string is finite) have been represented by dots in the
$(\Rhop_\chi$,$\Rhop_\psi)$-plane, and as can be seen in
Fig.~\ref{figuiposip}, for reasonable values of $\I$, there is no
available quantum state inside the tiny decreasing regions.
Consequently, the density line energy always grows with the fermions
densities and remains positive. Since the stationary curves of
$\widehat{U}$ are asymptotically proportional to $1/L_e < 1/L$ [see
Eq.~(\ref{energystat})], they coincide with the axis in the infinite
string limit. The surface describing $\widehat{U}(\Rhop_\chi,
\Rhop_\psi)$ has also been plotted in Fig.~\ref{figuiposip} in unit
normalized to $M^2$, and for minimal acceptable value of
$L_e=10/M$ just in order to show the influence of the finite length.


\paragraph{Tension.}

\begin{figure}
\begin{center}
\epsfig{file=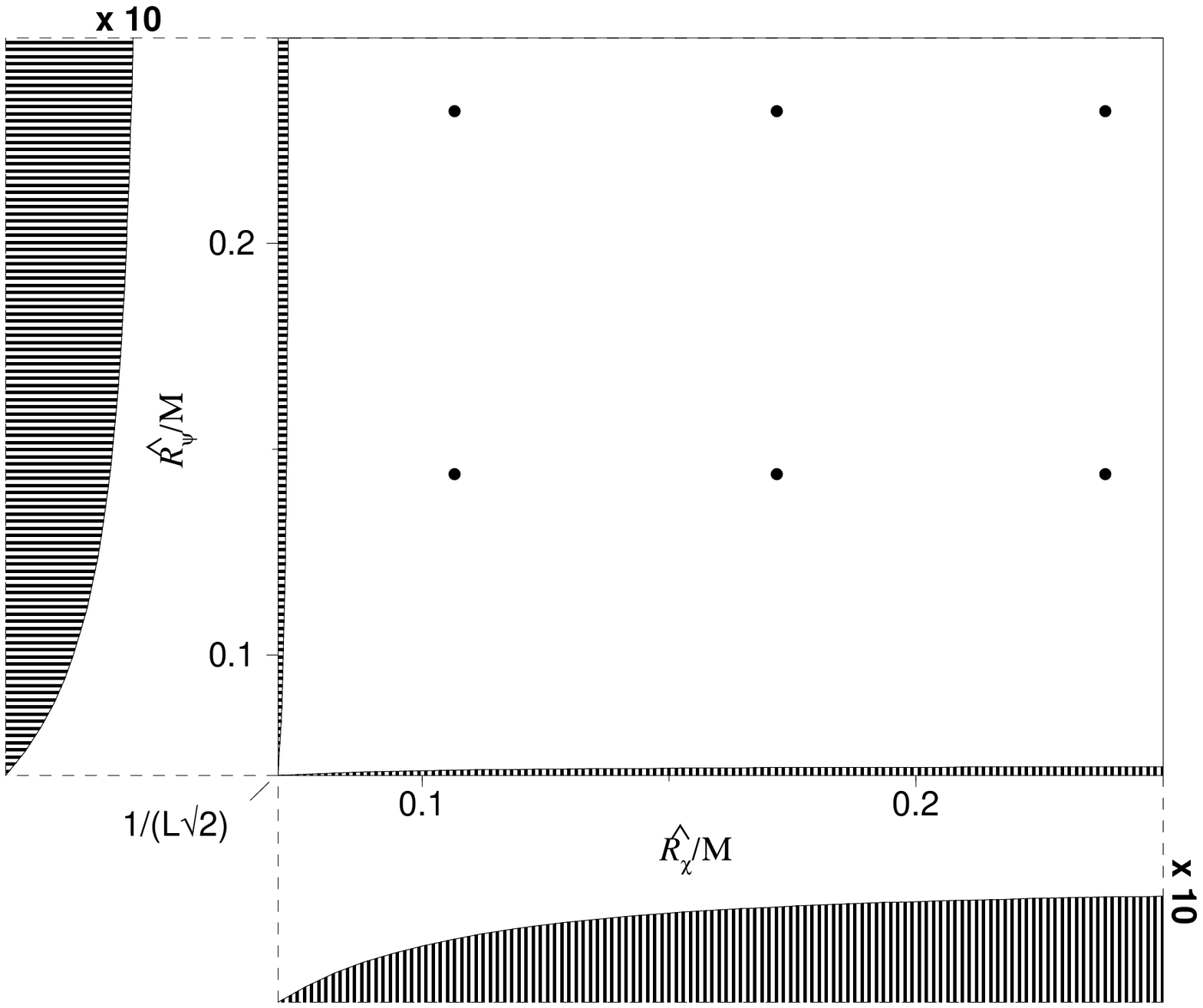,width=7cm}
\epsfig{file=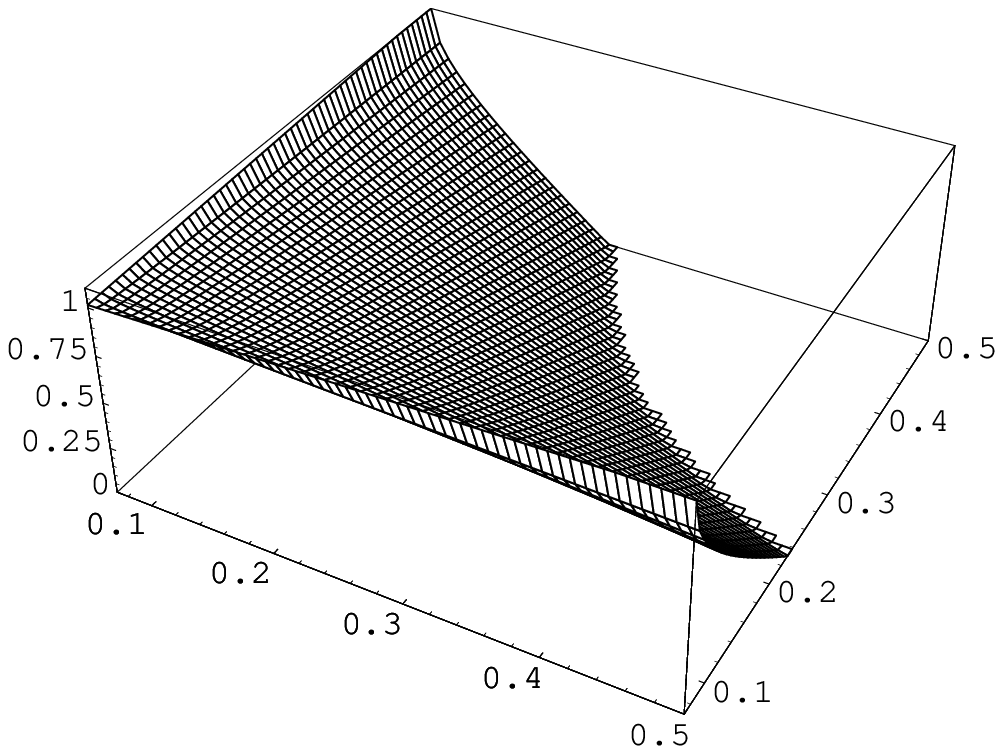,width=9cm}
\caption{Variation domains of the tension, plotted in unit of $M^2$,
for spacelike currents with $-4\pi<\I<0$, in the $(\Rhop_\chi/M,
\Rhop_\psi/M)$ plane. These regions have the same geometrical
properties as the line density energy ones in Fig.~\ref{figuiposip},
but this time, the zones with vertical hatches are domains where
$\widehat{T}$ could grow with respect to $\Rhop_\chi$, whereas the
horizontal ones correspond to growth with respect to $\Rhop_\psi$.
For reasonable values of $\I$, the discrete values of $\Rhop$,
represented by dots, are out of these regions, and the tension always
decrease with both fermion densities.}
\label{figtinegip}
\end{center}
\end{figure}

The study of the tension with respect to the fermion densities is
performed in the same way. As before the stationary curves
of $\widehat{T}$ with respect to $\Rhop_\chi$ or $\Rhop_\psi$ are
found from Eq.~(\ref{tfull}), and follow the same equation as those of
the energy per unit length in Eq.~(\ref{energystat}), although the
variation domains are not the same and have been plotted in
Fig.~\ref{figtinegip} for different values of the function $\I$.

\begin{figure}
\begin{center}
\epsfig{file=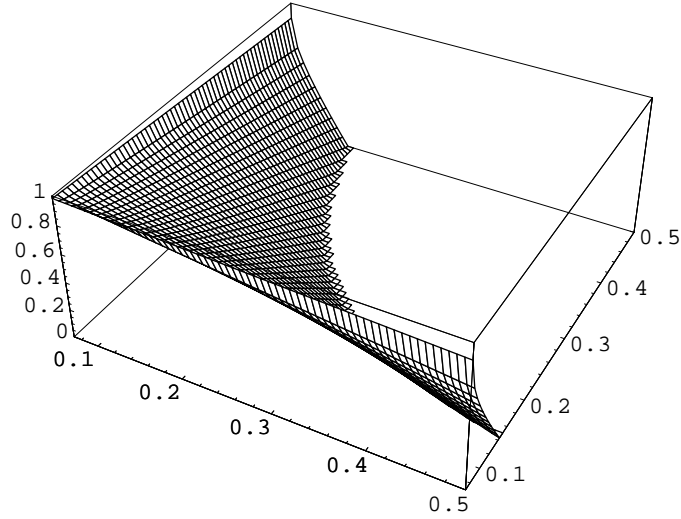,width=9cm}
\caption{The tension for timelike currents with $\I>0$, plotted in unit
of $M^2$ in the $(\Rhop_\chi/M,\Rhop_\psi/M)$ plane. Note the shift
near the axes due to the finite length of the string.}
\label{figtipos}
\end{center}
\end{figure}

\begin{figure}
\begin{center}
\epsfig{file=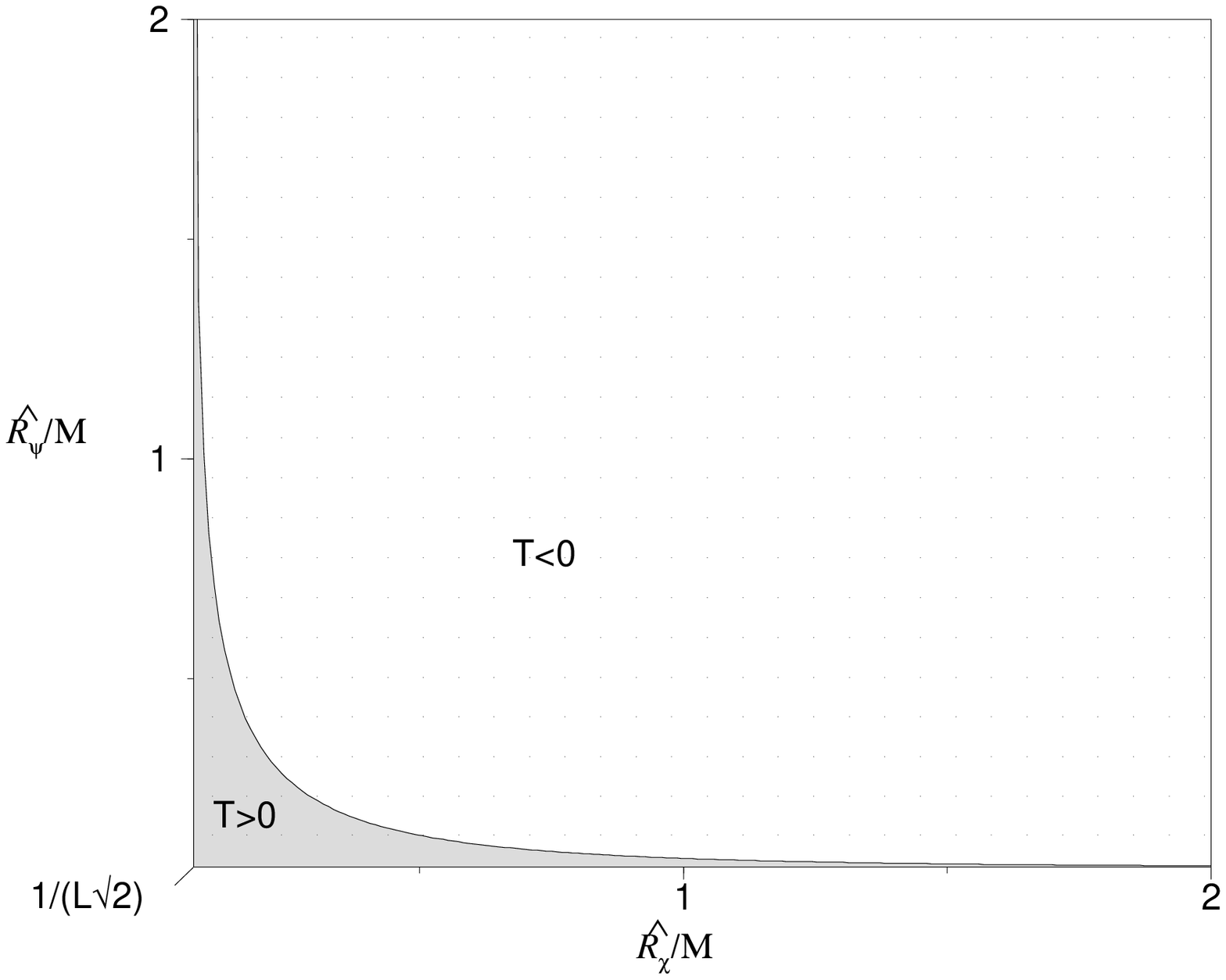,width=7cm}
\epsfig{file=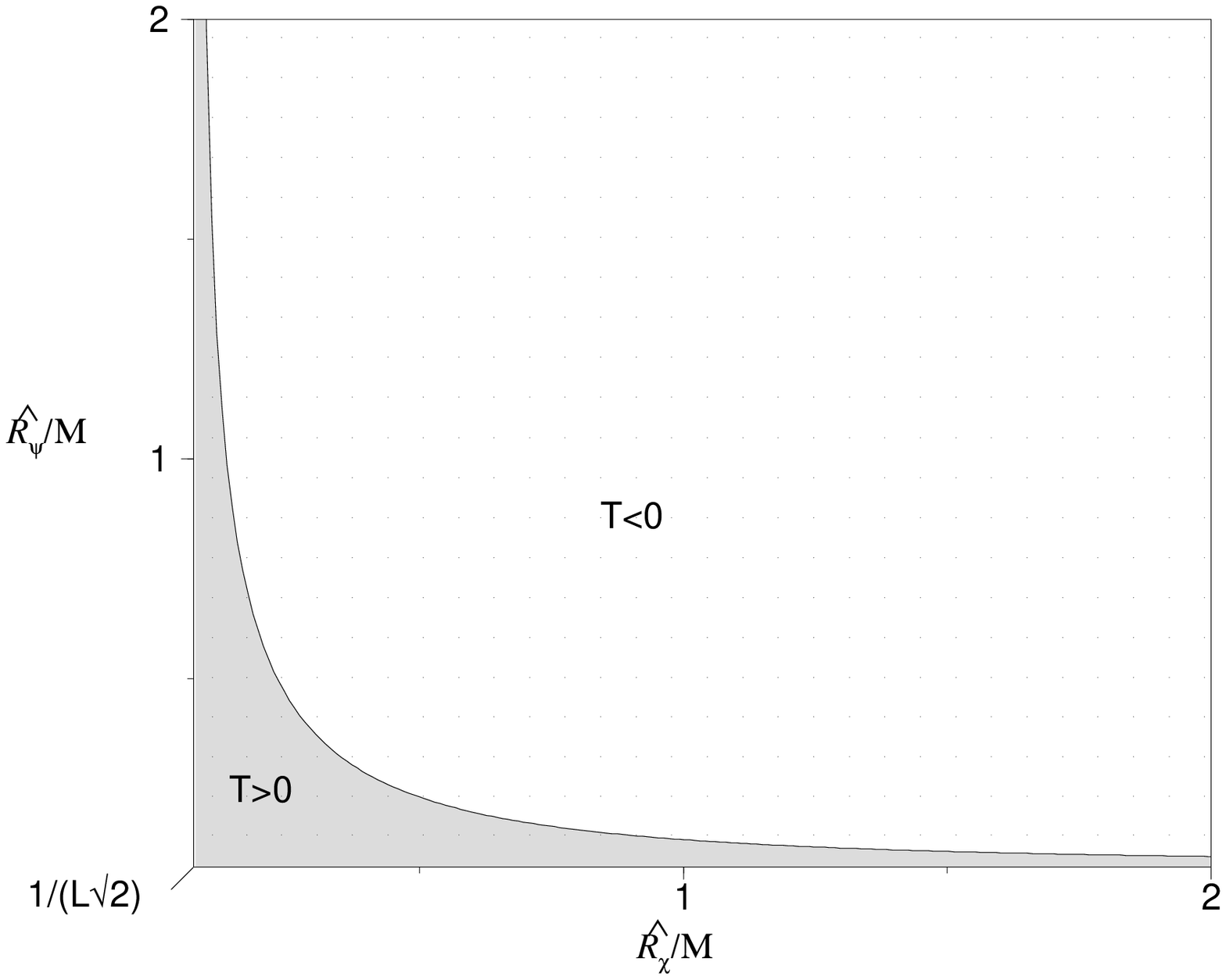,width=7cm} \caption{Sign of the
tension for timelike currents with $\I>0$, and spacelike currents
with $-4\pi<\I<0$, the curves have been plotted with $L_e=10/M$ in
aim at clearly separating the various regions. In the timelike
case, the domain where the tension is positive is closed, the
null tension curves intersecting the axes at
$\Rhop=\sqrt{2}L_eM^2/\I$, whereas for spacelike currents, the null
tension curves tends asymptotically to
$\Rhop=(2\pi/L_e)\sqrt{2/(16\pi^2-\I^2)}$. Recall that the regions
where $T<0$ are unstable with respect to transverse string
perturbations according to the macroscopic formalism.}
\label{figtnull}
\end{center}
\end{figure}

For timelike fermionic gauge current, $\I>0$, the tension
decreases faster than in the zeroth order case, with the damped
fermionic densities $\Rhop_\chi$ and $\Rhop_\psi$ as on
Fig.~\ref{figtipos}, and reaches negative values at large densities
(see Fig.~\ref{figtnull}). The backreaction just increases the
slope of the surface, and thus, the negative values are reached
more rapidly. As for the energy per unit length, the equivalent length
was chosen equal to $L_e=10/M$ in the following figures.

For spacelike fermionic gauge currents, $-4\pi<\I<0$, the backreaction
damps the decrease of the tension with respect to the damped fermion
densities. There are also tiny regions near the axis, with areas
inversely proportional to $L_e$, and where $\widehat{T}$ could grow with
respect to one of the state parameters $\Rhop_\chi$ or $\Rhop_\psi$
(see Fig.~\ref{figtinegip}). As previously, for reasonable values of
$\I$, the first discrete values of the parameters are out of these
domains, and the tension always decreases with both fermion densities.
Finally, the tension reaches negative values at large damped
fermion densities (see Fig.~\ref{figtnull}).

\subsubsection{Relevant values of the parameters}
\label{relevant}

The previous derivation of the backreaction is built on the
classical vortex background and it is acceptable only if the
backreacted gauge fields do not perturb appreciably the Higgs and
orthoradial gauge fields profiles (see Fig.~\ref{figback}). From
Eq.~(\ref{higgsmvt}), it will be the case only if $Q^t Q_t$ and $Q^z
Q_z$ can be neglected compared to $Q^\theta Q_\theta$. From
Eq.~(\ref{newgaugetilde}), and $Q_\theta Q^\theta \sim  m_b^2$, this
condition reads
\begin{eqnarray}
\frac{(q \cphi)^2 \Delta B \Sigma B}{Q_\theta Q^\theta} \sim
\frac{\Rho_\chi \Rho_\psi \sin{\left(\Theta_\chi-\frac{\pi}{4} \right)}
\sin{\left(\Theta_\psi -\frac{\pi}{4} \right)}}{\pi \eta^2}
\frac{m_b^2}{\pi \eta^2} \Delta \tilde{Q} \Sigma \tilde{Q} & \ll & 1,
\end{eqnarray}
or as function of $\I(\Theta_\psi,\Theta_\chi)$ and the damped fermion
densities,
\begin{eqnarray}
\label{domainok}
\I(\Theta_\chi,\Theta_\psi) \frac{\Rhop_\chi \Rhop_\psi}{\pi \eta^2}
& \ll & 1.
\end{eqnarray}
This condition is satisfied for damped fermion densities small
compared to the string energy scale, or for tiny values of the
function $\I(\Theta_\psi,\Theta_\chi)$.
Moreover, the backreacted gauge fields need to be small in
order to no perturb significantly the zero modes. From the equations of
motion (\ref{psimvt}) and (\ref{chimvt}), this condition leads to
$\Delta B, \Sigma B \ll \Rho$ and from Eq.~(\ref{newgaugetilde}) to
\begin{eqnarray}
\label{devlimok}
\frac{m_b^2}{\eta^2} \frac{c_{\psi(\chi)}}{\cphi} \Sigma (\Delta)
\tilde{Q}(0) \ll 1.
\end{eqnarray}
On the other hand, the maximum value of $\I$ in Eq.~(\ref{ifonction})
is clearly obtained when there are only particles or anti-particles
trapped in the string ($\Theta=0$ or $\Theta=\pi$), and deriving the
order of magnitude of the numerical integral in Eq.~(\ref{numintegspin}),
using equations (\ref{newgaugemvt}) and (\ref{fchi})-(\ref{fpsi}), one
shows that the large values of $\I$ (as $\I>4\pi$) can only be obtained
for model parameters which verify
\begin{equation}
\frac{m_b^2}{\eta^2} \frac{c_{\psi(\chi)}}{\cphi} \Sigma (\Delta)
\tilde{Q}(0) > 1.
\end{equation}
As a result, in order for the backreacted gauge fields not to modify
the equations of motion of the fermions at first order, the function
$\I$ has to be much smaller than $4\pi$. If it is not the case, then
the previous zero modes are no longer valid solutions and the
relevant equations of motion, in the case of the $\Psi$ fermions,
now read, from Eq.~(\ref{psimvt})
\begin{equation}
\label{pertzero}
\left\{
\begin{array}{lll}
\vspace{4pt}
\displaystyle
\left(\frac{\ud\xi_1}{\ud r} + \frac{1}{r}\left[-q \cpsir B(r) + m_1\right]
\xi_1(r) \right) - i g \varphi(r) 
\xi_4(r) & = & i [\mp(k+\omega) - q\cpsir \Delta B] \xi_2(r), \\
\vspace{4pt}
\displaystyle
\left(\frac{\ud\xi_2}{\ud r} + \frac{1}{r}\left[q \cpsir B(r) - m_2\right]
\xi_2(r) \right) - i g \varphi(r)
\xi_3(r) & = & i [\pm(k-\omega) - q\cpsir \Sigma B ] \xi_1(r), \\
\vspace{4pt}
\displaystyle
\left(\frac{\ud\xi_3}{\ud r} + \frac{1}{r}\left[-q \cpsil B(r) + m_3\right]
\xi_3(r) \right) + i g \varphi(r) 
\xi_2(r) & = & i [\mp(k-\omega) - q\cpsil \Sigma B] \xi_4(r), \\
\displaystyle
\left(\frac{\ud\xi_4}{\ud r} + \frac{1}{r}\left[q \cpsil B(r) - m_4\right]
\xi_4(r) \right)  + i g \varphi(r) 
\xi_1(r) & = & i [\pm(k+\omega) - q\cpsil \Delta B] \xi_3(r),
\end{array}
\right.
\end{equation}
where the angular dependence has not been written owing to
Eq.~(\ref{angseparation}) and assuming $m_1=m_3+n$.
The zero modes seem to acquire an effective mass proportional to
$\Delta B$ or $\Sigma B$. More precisely, they are no longer
eigenstates of the $\gamma^0 \gamma^3$ operator since new spinor
components appear [$\xi_1$, $\xi_4$ here, see Eq.~(\ref{zeromodes})
and Eq.~(\ref{antizeromodes})].
It is clearly a second order effect since the gauge coupling constant
$q$ can be removed in the previous equations (\ref{pertzero}) using
Eq.~(\ref{newgaugetilde}) and assuming
\begin{equation}
\xi=X+q^2\delta\xi,
\end{equation}
with $X$ the zero mode solution $k=-\omega$ for $\Psi$ fermions
[see Eq.~(\ref{psisolutions})], and $q^2\delta\xi$ the perturbation
induced by the backreaction~\cite{prep2}. As a result, for strong
backreaction, the semi-classical approach can no longer be
used, since such second order effects appear as the semi-classical
manifestations of the one loop quantum corrections, and thus, only a
full quantum theory would be well defined.
However, if there is only one kind of fermion trapped in the
string, $\Psi$ say, the zero modes are not affected by the
backreaction since $\Delta B$ is only generated from the $\Chi$
current, and therefore vanishes [see Eq.~(\ref{newgaugemvt})], so
$\xi=X$ is always solution of the equations of motion
(\ref{pertzero}), and identically for the $\Chi$ zero modes
alone~\cite{moreno}.
Note, that there is no contradiction with the usual index
theorem since it is derived for Dirac operators, and thus without
backreacted fields. This just shows that the modes propagating in
the vortex with strong backreacted gauge fields are no longer well
described by the usual zero modes.
Physically, it might be the signature of a tunneling of the zero modes
to another states. The massive modes which have not been considered
here could be more relevant in such cases.

On the other hand, the shape of the string might allow the fermion
densities $\Rhop_\chi$ and $\Rhop_\psi$ to reach the tiny regions
where the energy decreases with one of them (see
Fig.~\ref{figuiposip}), by means of the zero mode vacuum quantum
effects. The present toy model does not involve the effect of the
radius of curvature $R$ of the string, and it is reasonable that the
contribution of the zero mode vacuum to the energy per unit length
involves $R$ through a redefinition of $L_e$. If $L_e$ becomes
smaller than $L$, the first discrete values of the fermion densities
could be inside the hatched regions in Fig.~\ref{figuiposip}, since
the discrete values of the fermion densities only depend on the
physical length of the string $L$. Note that it would therefore be
necessary that the zero mode vacuum energy is negative, which is not
the case without curvature in the simple framework of
section \ref{toyvac}. Such effects could be relevant for vorton
stability, as, for a small radius of curvature, the string could
become unstable to fermion condensation.

Finally, the model can be used only at the tree order, and the
conditions (\ref{domainok}) and (\ref{devlimok}) are the validity
criteria of the above derivations.

\section{Comparison with the scalar case}

Owing to the fermionic two-dimensional quantization along the
string, the energy per unit length and the tension of a string
carrying massless fermionic currents have been derived up to the
first order in backreaction corrections. The state of the string
is found to be well defined with four state parameters which are
the densities of each fermion trapped in the string, and asymmetry
angles between particles and anti-particles in each fermion
family. It seems quite different than the bosonic charge carriers
case, where the current magnitude is the only relevant state
parameter~\cite{neutral}, however, this is the result of the
allowed purely classical approach where the superposition of many
quantum states can be view as only one classical state owing to
the bosonic nature of the charge carriers. As a result, there is a
degeneracy between the number of bosons trapped in the string and
the charge current. The quantization introduced to deal with
fermions naturally leads to separate the charge current from the
particle current through the existence of anti-particle
exitations. Moreover, the magnitude of the current can only
modify the equation of state at non-zeroth order because the
chiral nature of fermions trapped in the string requires
simultaneous exitations between the two families to lead to
non-lightlike charge currents.

Nevertheless, some global comparisons can be made with the scalar
case. First, for reasonable values of $\I \ll 4\pi$, the energy per
unit length grows with the fermionic densities, whereas the tension
decreases with them.  However, note the relevant parameter for the
change in behaviors of the tension and line density energy is the
function $\I(\Theta_\chi,\Theta_\psi)$ instead of the current
magnitude in the scalar case. As it was said, $\I$ quantifies, through
the asymmetry between the number of particles and anti-particles trapped
in the string, the efficiency of the charge current per particle to be
timelike or spacelike. The more positive is $\I$, the more timelike the
fermionic charge current per particle will be, and conversely the more
negative $\I$ is, the more spacelike it will be. Once again, this
difference with the scalar case appears as a result of the degeneracy
breaking between particle current and charge current due to the
fermionic nature of the charge carriers.

The stability of the string with respect to transverse
perturbations is given from the macroscopic formalism by the sign
of the tension (for line density energy positive)~\cite{formal},
and we find that instabilities always occur for densities roughly
close to $M/\sqrt{4\pi+\I}$, in finite domain for timelike
current, in infinite one for spacelike currents with $-4\pi<\I<0$.
Another new results are obtained from the multi-dimensional
properties of the equation of state, in particular the problem of
stability with respect to longitudinal perturbations differs from
the scalar barotropic case where the longitudinal perturbations
propagation speed is given by $c_L^2= -dT/dU$~\cite{formal}, and
therefore its two-dimensional form has to be derived to conclude
on these kinds of instabilities. Nevertheless, by analogy with
the scalar case, since, in the non-perturbed case and in the
infinite string limit, the equation of state verifies
$U+T=2M^2$~\cite{prep}, the longitudinal perturbation propagation
speed might be close to the speed of the light, even with small
backreaction, and therefore, only transverse stability would be
relevant in macroscopic string stability with massless fermionic
currents.

\section{Conclusion}

The energy per unit length and the tension of a cosmic string
carrying fermionic massless currents were derived in the frame of
the Witten model in the neutral limit. Contrary to bosonic charge
carriers, the two-dimensional quantization required to deal with
fermions, leads to more than one state parameter in order to
yield a well-defined equation of state. They can be chosen, at
zeroth order, as fermion densities trapped in the string
regardless of charge conjugation. The minimal backreaction
correction appears through the fermionic charge current magnitude
which involves the asymmetry angles between the number of
particles and anti-particles trapped in the string, and which
might be identified with the baryonic number of the plasma in
which the string was formed during the phase transition. As a
result, it is shown that fermionic charge currents can be
lightlike, spacelike as well as timelike. Moreover the line
energy density and the tension evolve globally as in the bosonic
charge carriers case, but it was found that the tension can take
negative values in extreme regions where the fermion densities are
close to the string mass, and where the string is therefore
unstable with respect to transverse perturbations according to
the macroscopic formalism.

The present model has been built on the generic existence of
fermionic zero modes in the string and follows only a
semi-classical approach. It is no longer valid for higher
corrections in the backreaction when they modify notably the vortex
background and seem to give effective mass to the previous zero
modes. It may be conjectured, at this stage, that in a full
quantum theory, the quantum loop corrections give mass to the
zero modes for high currents and consequently might lead to their
decay by the mean of massive states. Only chiral charge currents could
be stable on cosmic string carrying large fermionic massless currents
in such a case. Another possible effect, relevant for vortons stability,
may be expected for loops with small radius of curvature, by means of
the vacuum effects which could render the loop unstable to fermion
condensation.

It will be interesting to quantify such modifications on the equation of
state in future works, as the effects of worldsheet curvature, and the
modification of the density line energy and tension by the massive
bound states. The field of validity of the model could therefore be
extended to higher energy scales which would be more relevant for vortons
and string formation.

\acknowledgments I would like to thank P. Peter for many fruitful
discussions, and for his help to clarify the presentation. I also
wish to thank B. Carter who helped me to enlighten some aspect of
the subject.

\end{document}